\documentclass[sigconf, nonacm]{acmart}
\usepackage[linesnumbered,ruled,vlined]{algorithm2e}
\SetKwComment{Comment}{// }{}
\usepackage{amsmath}

\usepackage{enumitem}
\usepackage{multicol}
\usepackage{multirow}
\usepackage{float}
\usepackage{subfig}
\usepackage{balance}
\usepackage{bm}
\usepackage[normalem]{ulem}
\usepackage[table]{xcolor}
\usepackage{makecell}
\usepackage{tabularx}
\usepackage{booktabs}
\usepackage{array}
\usepackage{xcolor}
\usepackage{pifont}
\usepackage{bbding}
\usepackage{graphicx}
\usepackage{tcolorbox}
\usepackage{CJKutf8}
\usepackage{marginnote}
\usepackage{seqsplit}

\usepackage{framed}
\usepackage[dvipsnames, table]{xcolor} 
\definecolor{shadecolor}{gray}{0.95} 
\definecolor{ANNBlue}{HTML}{1F77B4}     
\definecolor{ANNTeal}{HTML}{17BECF}     
\definecolor{ANNOrange}{HTML}{FF7F0E}   
\definecolor{ANNGreen}{HTML}{2CA02C}    
\definecolor{ANNPurple}{HTML}{9467BD}   
\definecolor{ANNGrey}{HTML}{4D4D4D}     

\newcommand{\annsym}[1]{\scalebox{1.1}{$#1$}}

\newcommand{\CV}{\textcolor{ANNBlue}{\annsym{\blacktriangle}}}      
\newcommand{\EPV}{\textcolor{ANNTeal}{\annsym{\triangleright}}}     
\newcommand{\CA}{\textcolor{ANNOrange}{\annsym{\bigstar}}}          
\newcommand{\SEG}{\textcolor{ANNGreen}{\annsym{\blacksquare}}}      
\newcommand{\GIS}{\textcolor{ANNPurple}{\annsym{\bullet}}}          
\newcommand{\FV}{\textcolor{ANNGrey}{\annsym{\blacklozenge}}}       

\newcommand{\sym}[2][1.25]{\raisebox{0.15ex}{\scalebox{#1}{#2}}}
\newcommand{\xmark}{\sym{\textcolor{red!70!black}{\ding{55}}}} 

\newcommand\vldbdoi{XX.XX/XXX.XX}
\newcommand\vldbpages{XXX-XXX}
\newcommand\vldbvolume{19}
\newcommand\vldbissue{13}
\newcommand\vldbyear{2026}
\newcommand\vldbauthors{\authors}
\newcommand\vldbtitle{\shorttitle} 
\newcommand\vldbavailabilityurl{https://github.com/OpenDataBox/Disk-ANN-Evaluation}
\newcommand\vldbpagestyle{empty} 

\begin{document}
\title{Disk-Resident Graph ANN Search: An Experimental Evaluation}

\author{Xiaoyu Chen}
\authornote{Equal contribution.}
\affiliation{
  \institution{Shanghai Jiao Tong Univ.}
  \country{}
}
\email{Chenxy1@sjtu.edu.cn}

\author{Jinxiu Qu}
\authornotemark[1]
\affiliation{
  \institution{Shanghai Jiao Tong Univ.}
  \country{}
}
\email{afuloowa@sjtu.edu.cn}

\author{Yitong Song}
\authornote{Corresponding author.}
\affiliation{
\institution{Hong Kong Baptist Univ.}\country{}
}
\email{ytsong@comp.hkbu.edu.hk}

\author{Shuhang Lu}
\affiliation{%
  \institution{Shanghai Jiao Tong Univ.}
  \country{}
}
\email{ts1989@sjtu.edu.cn}

\author{Huiling Li}
\affiliation{
  \institution{Hong Kong Baptist Univ.}
  \country{}
}
\email{cshlli@comp.hkbu.edu.hk}

\author{Minghui Jiang}
\affiliation{
  \institution{Shanghai Jiao Tong Univ.}
  \country{}
}
\email{tuanzi7@sjtu.edu.cn}

\author{Wei Zhou}
\affiliation{\institution{{Shanghai Jiao Tong Univ.}}\country{}}
\email{{weizhoudb@sjtu.edu.cn}}

\author{Jianliang Xu}
\affiliation{
  \institution{Hong Kong Baptist Univ.}
  \country{}
}
\email{xujl@comp.hkbu.edu.hk}

\author{Xuanhe Zhou}
\affiliation{\institution{{Shanghai Jiao Tong Univ.}}\country{}}
\email{{zhouxuanhe@sjtu.edu.cn}}

\author{Fan Wu}
\affiliation{
  \institution{Shanghai Jiao Tong Univ.}
  \country{}
}
\email{fwu@cs.sjtu.edu.cn}

\begin{abstract}
As data volumes grow while memory capacity remains limited, disk-resident graph-based approximate nearest neighbor (ANN) methods have become a practical alternative to memory-resident designs, shifting the bottleneck from computation to disk I/O. However, since their technical designs diverge widely across storage, layout, and execution paradigms, a systematic understanding of their fundamental performance trade-offs remains elusive.

This paper presents a comprehensive experimental study of disk-resident graph-based ANN methods. First, we decompose such systems into five key technical components, i.e., storage strategy, disk layout, cache management, query execution, and update mechanism, and build a unified taxonomy of existing designs across these components. 
Second, we conduct fine-grained evaluations of representative strategies for each technical component to analyze the trade-offs in throughput, recall, and resource utilization. 
Third, we perform comprehensive end-to-end experiments and parameter-sensitivity analyses to evaluate overall system performance under diverse configurations. 
Fourth, our study reveals several non-obvious findings: (1) vector dimensionality reshapes component effectiveness, necessitating dimension-aware design; (2) existing layout strategies exhibit surprisingly low I/O effectiveness; (3) page size strongly affects feasibility and efficiency, with smaller pages often performing better; and (4) update strategies involve workload-dependent trade-offs. 
Based on these findings, we provide practical design guidelines and identify promising research directions.
\end{abstract}

\maketitle

\pagestyle{\vldbpagestyle}
\begingroup\small\noindent\raggedright\textbf{PVLDB Reference Format:}\\
\vldbauthors. \vldbtitle. PVLDB, \vldbvolume(\vldbissue): \vldbpages, \vldbyear.\\
\href{https://doi.org/\vldbdoi}{doi:\vldbdoi}
\endgroup
\begingroup
\renewcommand\thefootnote{}\footnote{\noindent
This work is licensed under the Creative Commons BY-NC-ND 4.0 International License. Visit \url{https://creativecommons.org/licenses/by-nc-nd/4.0/} to view a copy of this license. For any use beyond those covered by this license, obtain permission by emailing \href{mailto:info@vldb.org}{info@vldb.org}. Copyright is held by the owner/author(s). Publication rights licensed to the VLDB Endowment. \\
\raggedright Proceedings of the VLDB Endowment, Vol. \vldbvolume, No. \vldbissue\ %
ISSN 2150-8097. \\
\href{https://doi.org/\vldbdoi}{doi:\vldbdoi} \\
}\addtocounter{footnote}{-1}\endgroup

\ifdefempty{\vldbavailabilityurl}{}{
\vspace{.3cm}
\begingroup\small\noindent\raggedright\textbf{PVLDB Artifact Availability:}\\
The source code, data, and/or other artifacts have been made available at \url{\vldbavailabilityurl}.
\endgroup
}

\section{Introduction}
\label{sec:intro}

Graph-based Approximate Nearest Neighbor (ANN) methods organize high-dimensional vectors as a proximity graph and answer queries through graph traversal, achieving strong search efficiency–accuracy trade-offs. Owing to their superior performance, they have become the state-of-the-art~\cite{ANNSurvey, ANNSurvey2} and are widely adopted in production systems~\cite{Milvus, elasticsearch, pgvector}.

\begin{figure}[!t]
    \centering
    \includegraphics[width=0.9\linewidth]{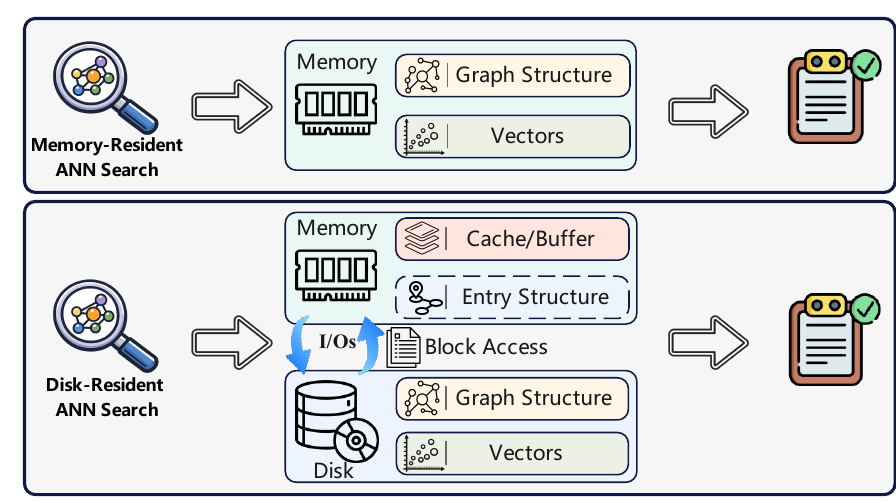}
    \vspace{-0.15in}
    \caption{Memory-resident vs. disk-resident ANN search.}
    \label{fig:ann-search-comparsion}
    \vspace{-0.2in}
\end{figure}

Early graph-based ANN methods~\cite{HNSW, NSW, NSG, tau-MNG, NSSG} are mainly designed to be memory-resident, with both raw vectors and graph structures stored entirely in memory. However, as data volumes grow and memory remains expensive, this design becomes impractical, motivating a large body of disk-resident graph-based methods~\cite{DiskANN, Starling, Gorgeous, PageANN, PipeANN, XN-Graph, DGAI, FreshDiskANN, ODINANN, AiSAQ, LM-DiskANN, BAMG, margo}. As shown in Figure~\ref{fig:ann-search-comparsion}, disk-resident designs differ fundamentally from memory-resident ones and require several additional design considerations.


\noindent
\textbf{(1) Heterogeneous Storage and Layout.} 
Disk-resident methods must determine how to place key data components, including raw vectors, graph index structures, and auxiliary data (e.g., entry structures and compressed vectors), across heterogeneous memory–disk storage. Different storage strategies directly determine both memory footprint and I/O volume, thereby exposing inherent trade-offs between query efficiency and memory usage~\cite{AiSAQ, DiskANN}. 
Moreover, since disk accesses are performed at block granularity, the physical layout of data within and across disk blocks plays a critical role in improving data locality and query performance~\cite{DiskANN, Starling, PageANN, Gorgeous}.

\noindent
\textbf{(2) Block-Level and Cache-Aware Query Processing.}
In disk settings, ANN search no longer accesses individual vectors but entire disk blocks, leading to access patterns that differ fundamentally from memory designs. 
As a result, the primary performance bottleneck shifts from in-memory computation to disk I/O.
To mitigate disk access latency, existing methods incorporate (i) cache-aware techniques~\cite{GoVector,XN-Graph,Gorgeous,PageANN}, and (ii) asynchronous execution~\cite{PipeANN,PageANN,Starling,DGAI} during graph traversal, aiming to reduce both the number of I/O operations and the time spent waiting for I/O.

\noindent
\textbf{(3) Higher Update Overhead.} 
Memory-resident methods naturally support efficient insertions, as updates can be applied directly to in-memory data structures. In contrast, insertions in disk-resident systems often incur high overhead because a single update may require modifying multiple disk blocks. Deletions are even more challenging, as removing vectors can disrupt graph connectivity and consequently degrade query accuracy.

\begin{table*}[!t]
\centering
\vspace{-0.1in}
\caption{Comparison of Disk-Resident Graph-Based ANN Methods Across Key Design Dimensions.}
\vspace{-0.15in}
\label{tab:compared_methods}
\small
\renewcommand{\arraystretch}{1.2} 
\resizebox{0.98\linewidth}{!}{
\begin{tabular}{l c c c c c c}
\toprule
\multirow{2}{*}{\textbf{Method}} & 
\multirow{2}{*}{\textbf{Storage Strategy}} & 
\multicolumn{2}{c}{\textbf{Layout Strategy}} & 
\multirow{2}{*}{\textbf{Cache Strategy}} & 
\multirow{2}{*}{\textbf{Asynchronous}} & 
\multirow{2}{*}{\textbf{Update Mechanism}} \\ 
\cmidrule(lr){3-4}
 &  & \textbf{Global} & \textbf{Local} &  &  &  \\ 
\midrule

\makecell[l]{DiskANN~\cite{DiskANN} / FreshDiskANN~\cite{FreshDiskANN}} & Major-in-Disk  & Coupled & ID-Based & Static & \xmark & Out-of-Place \\
\hline
AiSAQ~\cite{AiSAQ} & All-in-Disk & Coupled & ID-Based & \xmark & \xmark & \xmark \\
\hline
Starling~\cite{Starling} & Major-in-Disk  & Coupled & Heuristic-Based & Static & Compute-Driven & \xmark \\
\hline
Gorgeous~\cite{Gorgeous} & Major-in-Disk & Coupled & Graph-Replicated & Static & Compute-Driven & \xmark \\
\hline
PageANN~\cite{PageANN} & All-in-Disk \& Major-in-Disk & Coupled & Clustering-Based & Static & Compute-Driven & \xmark \\
\hline
\makecell[l]{PipeANN~\cite{PipeANN} / OdinANN~\cite{ODINANN}} & Major-in-Disk & Coupled & ID-Based & \xmark & I/O-Driven & In-Place \\
\hline
DGAI~\cite{DGAI} & Major-in-Disk & Decoupled & Clustering-Based & Hybrid & Compute-Driven & In-Place \\
\hline
AlayaLaser~\cite{alayalaser} & All-in-Disk & Coupled & ID-Based & Static & I/O-Driven &  \xmark \\
\bottomrule
\end{tabular}
}
\makebox[\linewidth][l]{\footnotesize \xmark\ denotes not supported or not specified.}
\vspace{-0.25in}
\end{table*}

\begin{figure}[!t]
    \centering
    \includegraphics[width=\linewidth]{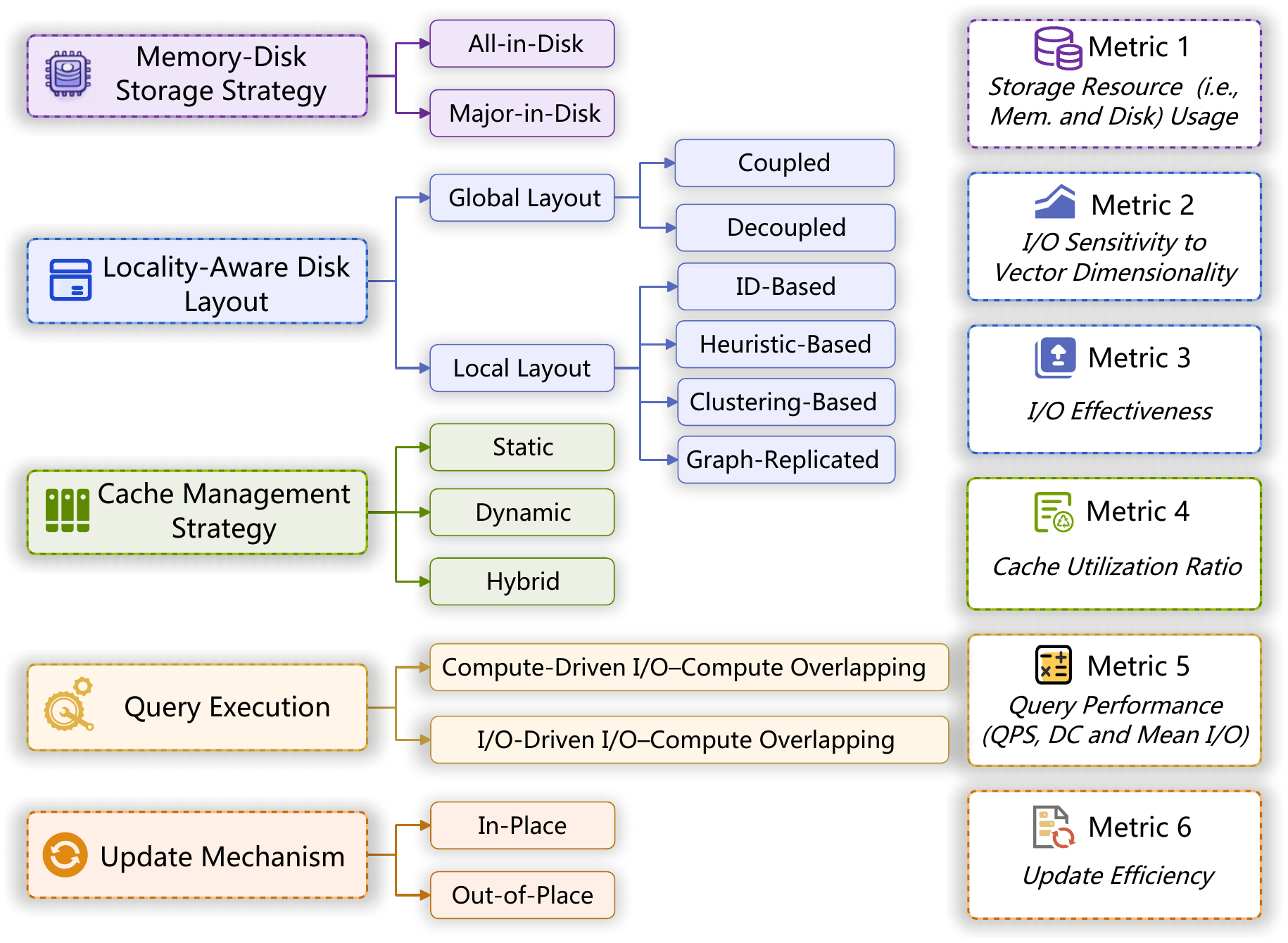}
    \vspace{-0.15in}
    \caption{Technology decomposition and experimental study.}
    \vspace{-0.25in}
    \label{fig:tech-decomposition}
\end{figure}

\subsection{Our Motivation}

\noindent
\textbf{(1) Lack of a Systematic Taxonomy and Comparison.} 
Prior studies~\cite{echihabi2020return, FANNExp, graphSurvey} primarily evaluate memory-resident ANN methods, while disk-resident approaches have recently evolved with substantially different designs. A recent study~\cite{DesignSpace} explores I/O optimization across three design dimensions, but does not provide a unified taxonomy or a comprehensive comparison of methods. 


\noindent
\textbf{(2) Absence of a Unified Evaluation Framework.}
Existing methods are often evaluated in isolation or under method-specific configurations (e.g., fixed 16KB page sizes), which prevents fair and consistent comparisons. Therefore, a unified evaluation framework for systematically assessing their performance remains absent.

\noindent
\textbf{(3) Lack of Practical Design Guidelines.} 
Disk-resident graph-based ANN systems consist of many interdependent components whose performance is jointly shaped by diverse data characteristics and system factors. However, practical guidelines for selecting appropriate design choices under varying data characteristics, system settings, and deployment constraints remain absent.

\subsection{Our Contributions}
\label{sec:our_contributions}
Motivated by this, we present an in-depth study of disk-resident graph-based ANN methods and make the following contributions.

\noindent
\textbf{(1) Technology Decomposition and Taxonomy (Section~\ref{sec:technique-overview}).}
As illustrated in Figure~\ref{fig:tech-decomposition}, we decompose existing methods into five core technique components: (i) memory–disk storage strategy, (ii) locality-aware disk layout, (iii) cache management strategy, (iv) query execution, and (v) update mechanism. For each component, we further establish a structured taxonomy by categorizing existing approaches according to their underlying design principles, and define corresponding evaluation metrics to enable principled comparison across different methods.

\noindent
\textbf{(2) Fine-Grained Evaluation of Technique Components (Section~\ref{sec:fine_performance}).} Leveraging carefully designed component-level metrics, we conduct controlled and fine-grained experiments on representative strategies within each technique component. We quantify their performance trade-offs, analyze sensitivity to key factors (e.g., dimensionality), and assess their impact on query efficiency.

\noindent
\textbf{(3) Overall End-to-End Performance Evaluation (Section~\ref{sec:overall_performance}).}
We further conduct comprehensive end-to-end evaluations under a unified and fair testbed. Our study includes a diverse set of representative systems (Table~\ref{tab:compared_methods}), each embodying different combinations of storage, layout, caching, asynchronous execution, and update strategies. We also systematically examine the impact of critical parameters (e.g., page size and beam width) on overall performance.

\noindent
\textbf{(4) Insightful Findings (Section~\ref{sec:experiments}).}
Based on our experimental results and in-depth analysis, we distill a set of insightful findings, including both new observations and cross-method validations of conclusions reported in prior work~\cite{DesignSpace,song2026low,alayalaser}:

\noindent
\ding{182} PageANN excels at low dimensionality (i.e., $d$), whereas PipeANN dominates in high $d$. In high $d$, poor block locality makes multi-block accesses unavoidable, rendering asynchronous execution critical for hiding access latency. In contrast, at low $d$, stronger block locality makes disk layout more important for reducing accesses. These results highlight the need for \emph{dimension-aware design choices}.

\noindent
\ding{183} As $d$ and required recall increase, the index–vector decoupled layout becomes increasingly advantageous over the widely used coupled layout. Moreover, overall I/O effectiveness remains consistently \emph{low} across all layout strategies, typically below 15\%.

\noindent
\ding{184} Asynchronous execution masks computation costs, making I/O dominant in search latency. However, as $d$ increases, computation overhead becomes more significant.

\noindent
\ding{185} Lower $d$ favors \emph{dynamic/hybrid} caching, while higher $d$ favors \emph{static (graph-prioritized)} caching.

\noindent
\ding{186} With small page sizes, carefully selecting high-quality neighbors is critical, whereas with larger page sizes, retaining more neighbors becomes advantageous.

\noindent
\ding{187} Some methods become infeasible under small page sizes since a single page cannot accommodate their required data structures. When feasible, \emph{smaller page sizes} generally yield higher efficiency.

\noindent
\ding{188} In-place updates favor query-heavy workloads, whereas out-of-place updates better suit update-heavy or balanced workloads. 

\noindent
\textbf{(5) Usage Guidelines and Research Directions (Section~\ref{sec:lessons}).}
We summarize practical guidelines for selecting and configuring disk-resident graph-based ANN techniques, and outline several promising directions for future research.
\section{Preliminaries}
\label{sec:preliminary}

\noindent
\textbf{ANN Search.}
Given a vector dataset $\mathcal{D}=\{\mathbf{v}_1,\dots,\mathbf{v}_N\}$, a query vector $\mathbf{q}$, and a similarity metric $sim(.,.)$, ANN search retrieves a set $\mathcal{S}'\subseteq\mathcal{D}$ of $k$ vectors that are approximately most similar to $\mathbf{q}$.
We focus on approximate search rather than exact search, as the latter is prohibitively expensive in high-dimensional spaces~\cite{HNSW,NSG}.
By relaxing exactness, ANN achieves substantially higher efficiency, with accuracy commonly measured by $\mathrm{Recall@}k= |\mathcal{S}'\cap\mathcal{S}|/{k}$,
where $\mathcal{S}$ denotes the exact $k$NN results.
The goal of ANN search is to maximize recall (often $r\!\ge\!0.9$) while minimizing query latency.



\noindent
\textbf{Disk-Resident Graph-Based ANN Search.}
Graph-based methods~\cite{HNSW,NSG,NSSG,NSW,LSH-APG,tau-MNG} achieve an excellent efficiency--accuracy trade-off by organizing vectors into a proximity graph and answering queries via graph traversal~\cite{ANNSurvey,ANNSurvey2}.
However, as data volume grows, storing both high-dimensional vectors and graph adjacency structures entirely in memory becomes impractical.
This has motivated a class of \emph{disk-resident graph-based ANN} methods~\cite{alayalaser, DiskANN, Starling, Gorgeous, PageANN, PipeANN, XN-Graph, DGAI, TRIM, FreshDiskANN, ODINANN, AiSAQ, LM-DiskANN, BAMG, LSM-VEC, margo, song2026disk,song2026vector,tenginedb}, which place memory-intensive components on disk while retaining lightweight structures in memory to preserve efficient traversal.

\begin{figure}[!t]
    \centering
    \includegraphics[width=\linewidth]{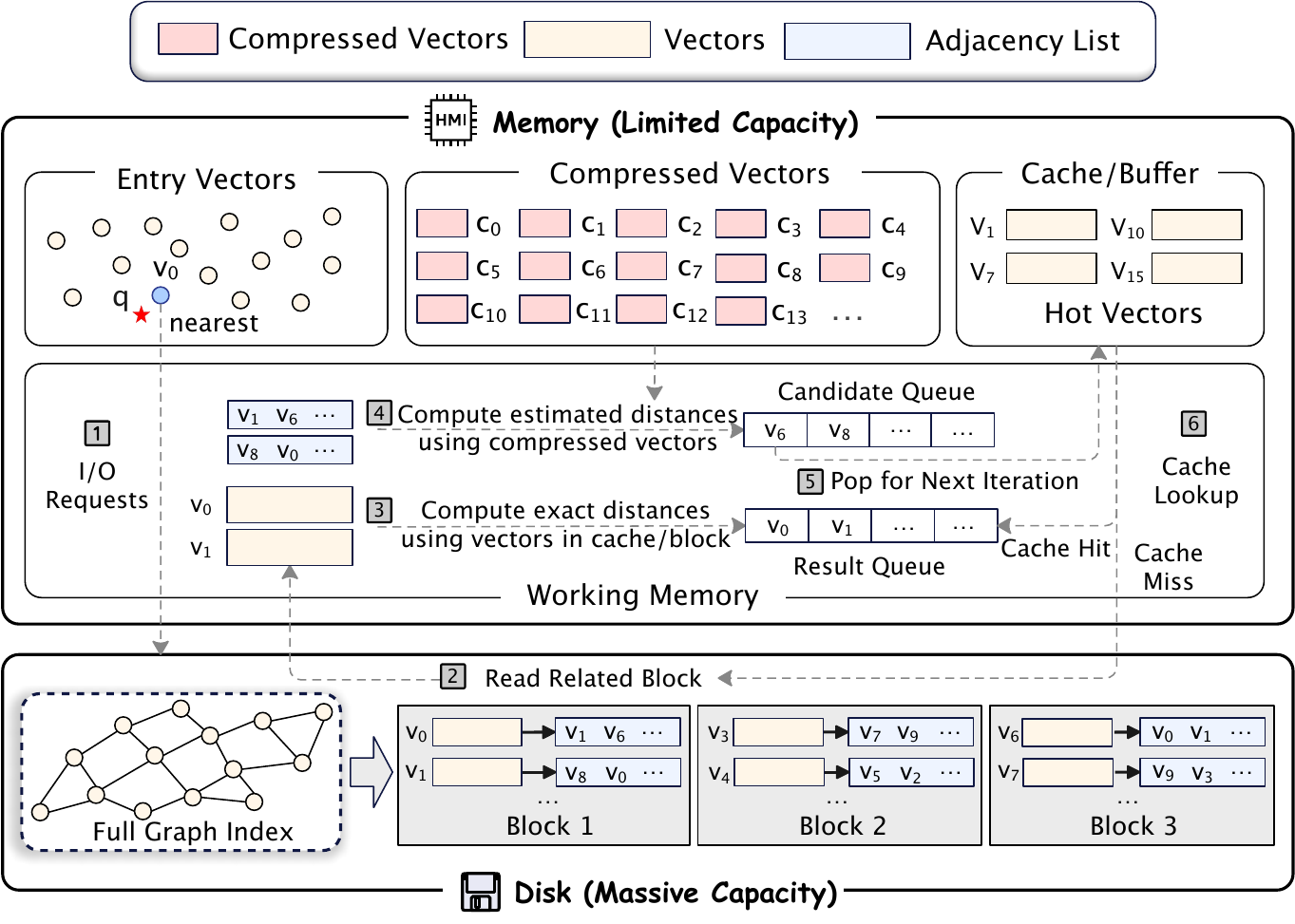}
   \vspace{-0.2in}
    \caption{Disk-resident graph-based ANN search.}
   \vspace{-0.15in}
    \label{fig:case}
\end{figure}

Figure~\ref{fig:case} illustrates a representative disk-resident graph-based ANN workflow, as adopted by DiskANN~\cite{DiskANN}.
In memory, three lightweight components are maintained:
(1) \emph{compressed vectors} (e.g., Product Quantization (PQ) codes~\cite{pq}) for fast distance estimation,
(2) \emph{entry vectors} for traversal initialization, and
(3) a \emph{fixed-size buffer} that caches frequently accessed disk blocks.
In contrast, storage-intensive data, including
(1) full \emph{graph adjacency lists} and
(2) \emph{raw vectors}, are stored on disk.
During ANN search, working memory maintains a \emph{candidate queue} of unvisited vectors ordered by estimated distance and a \emph{result queue} of the current top-$k$ vectors ranked by exact distance. The search starts from an entry vector, whose disk block is fetched to compute exact distances and expand its neighbors using estimates from in-memory compressed vectors. It then repeatedly expands the nearest unvisited candidate by accessing its disk block (or cache if available) until the smallest estimated candidate distance exceeds the worst exact distance in the result queue. The current top-$k$ vectors are then returned.

\vspace{-0.1in}
\section{Technology Analysis}
\label{sec:technique-overview}

The performance of disk-resident graph-based ANN search depends on multiple interacting design components that jointly affect query efficiency, recall, and system resource utilization. We decompose these components into several fundamental building blocks.

\noindent
\textbf{(1) Memory-Disk Storage Strategy.}
Disk-resident ANN systems must determine how key components, including compressed vectors, graph adjacency lists, and raw vectors, are kept between memory and disk.
Different storage strategies directly affect memory footprint and disk I/O volume, thereby inducing inherent trade-offs between query performance and resource utilization.


\noindent
\textbf{(2) Locality-Aware Disk Layout.} 
Beyond deciding which components reside on disk, disk-resident ANN systems must also organize how disk-resident data are laid out.
The physical layout directly affects data locality and I/O efficiency.
Therefore, layout optimization has attracted significant research attention~\cite{Starling,margo,Gorgeous,PageANN,DiskANN++,DGAI,TRIM}.


\noindent
\textbf{(3) Cache Management Strategy.}
Disk-resident ANN systems rely on caching to reduce disk I/O by retaining frequently accessed data in memory.
The effectiveness of caching depends on both \emph{what} data are cached and \emph{when} caching decisions are made, leading to different performance in terms of cache hit ratio and I/O reduction.

\noindent
\textbf{(4) Query Execution.}
Although most graph-based ANN methods follow a similar graph traversal paradigm, execution-level optimizations are critical to query efficiency. In particular, asynchronous I/O and prefetching overlap disk access with computation to reduce query latency, introducing trade-offs between query efficiency and system resource utilization.

\noindent
\textbf{(5) Update Mechanism.}
To support dynamic workloads, disk-resident ANN systems adopt different update mechanisms~\cite{ODINANN,FreshDiskANN}.
These mechanisms introduce distinct trade-offs among update efficiency, query performance, and maintenance overhead.

\subsection{Memory-Disk Storage Strategy}
Disk-resident graph-based ANN systems must carefully place three main components across memory and disk: (1) \emph{compressed vectors} (typically PQ codes), (2) \emph{graph adjacency lists}, and (3) \emph{raw vectors}. These components scale linearly with data size, whereas other structures, such as entry vectors/graph and caches, incur only bounded overhead. Based on whether scale-growing data is stored in memory, existing designs can be broadly classified into \emph{major-in-disk} and \emph{all-in-disk} strategies. Table~\ref{tab:mem-disk-summary} summarizes representative methods under these two categories.



\begin{table}[t]
\centering
\scriptsize
\setlength{\tabcolsep}{2.2pt}
\renewcommand{\arraystretch}{1.3}
\setlength{\extrarowheight}{1.3pt}

\caption{Memory--disk storage strategies. \textit{Legend:} \EPV\ = Entry Vectors, \CV\ = Compressed Vectors, 
\SEG\ = Sampled Entry Graph, \CA\ = Cache, \GIS\ = Adjacency Lists, and \FV\ = Vectors.}
\vspace{-0.15in}
\label{tab:mem-disk-summary}

\begin{tabularx}{\columnwidth}{|>{\centering\arraybackslash}m{0.9cm}|l| >{\centering\arraybackslash}m{0.9cm}| >{\centering\arraybackslash}m{0.9cm}|X|}
\hline
\textbf{Type} & \textbf{Method} & \textbf{Memory} & \textbf{Disk} & \textbf{Applications} \\
\hline

\multirow{9}{*}{\parbox[c]{0.9cm}{\centering Major-\\in-Disk$^{\star}$}} &
DiskANN~\cite{DiskANN} & \CV\ \CA & \GIS\ \FV &
\multirow{9}{=}{\parbox[t]{\hsize}{\raggedright\arraybackslash
\par\noindent(1) Low-latency/High-QPS ANN services.
\par\noindent(2) Deployments with relatively sufficient memory budget.}} \\
\cline{2-4}
& Starling~\cite{Starling}  & \CV\ \SEG\ \CA & \GIS\ \FV & \\
\cline{2-4}
& Gorgeous~\cite{Gorgeous}  & \CV\ \SEG\ \CA & \GIS\ \FV & \\
\cline{2-4}
& MARGO~\cite{margo}        & \CV\ \SEG\ \CA & \GIS\ \FV & \\
\cline{2-4}
& BAMG~\cite{BAMG}          & \CV\ \SEG      & \GIS\ \FV & \\
\cline{2-4}
& XN-Graph~\cite{XN-Graph}  & \CV\ \SEG      & \GIS\ \FV & \\
\cline{2-4}
& DGAI~\cite{DGAI}          & \CV\ \CA       & \GIS\ \FV & \\
\cline{2-4}
& PipeANN~\cite{PipeANN}    & \CV\ \SEG      & \GIS\ \FV & \\
\cline{2-4}

& \multirow{2}{*}{PageANN~\cite{PageANN}} &
  \multirow{2}{*}{\CV$^{\dagger}$ \SEG\ \CA} &
  \multirow{2}{*}{\CV$^{\dagger}$ \GIS\ \FV} & \\
\cline{1-1}\cline{5-5}

\multirow{4}{*}{\parbox[c]{0.9cm}{\centering All-in-\\Disk$^{\diamond}$}} &
& & &
\multirow{4}{=}{\parbox[t]{\hsize}{\raggedright\arraybackslash
\par\noindent(1) Memory-tight deployments (edge/on-device).
\par\noindent(2) Scale-sensitive applications where dataset-scale \CV\ cannot fit in memory.}} \\
\cline{2-4}
& AiSAQ~\cite{AiSAQ}             & \EPV & \CV\ \GIS\ \FV & \\
\cline{2-4}
& LM-DiskANN~\cite{LM-DiskANN}   & \CA  & \CV\ \GIS\ \FV & \\
\cline{2-4}
& LSM-VEC~\cite{LSM-VEC}         & \SEG\ \CA & \GIS\ \FV & \\
  \cline{2-4}
  & AlayaLaser~\cite{alayalaser}   & \EPV \CA  & \CV\ \GIS\ \FV & \\
\hline
\end{tabularx}

\vspace{2pt}
\begin{minipage}{\columnwidth}
\footnotesize
$^{\star}$ \textbf{Major-in-Disk}: compressed vectors in memory; adjacency lists and vectors on disk.\par
$^{\diamond}$ \textbf{All-in-Disk} (memory is still utilized): only auxiliary data in memory; all scale-growing components (i.e., compressed/raw vectors, and adjacency lists) on disk. \par
$^{\dagger}$ Budget-dependent placement.
\end{minipage}
\vspace{-0.1in}
\end{table}

\noindent
\textbf{Major-in-Disk.}
Major-in-disk is the most widely adopted storage strategy~\cite{DiskANN,Starling,Gorgeous,margo,BAMG,XN-Graph,DGAI,PipeANN}, where compressed vectors reside in memory and graph adjacency lists together with raw vectors are stored on disk. 
The key insight is that compressed representations are lightweight yet frequently accessed during graph traversal for neighbor distance estimation, making them well suited for memory residency. 
As a result, major-in-disk designs enable faster candidate scoring and prioritization with fewer disk accesses, substantially reducing disk I/O during search.
However, this benefit comes at the cost of memory footprints that grow linearly with data scale and may become prohibitive under strict memory budgets.

\noindent
\textbf{All-in-Disk.} 
All-in-disk methods ~\cite{alayalaser,AiSAQ,LM-DiskANN,LSM-VEC,PageANN} place all scale-growing components on disk, keeping only a small and controllable set of auxiliary structures (e.g., entries and caches) in memory.
This design minimizes memory footprint and makes all-in-disk designs particularly suitable for memory-constrained environments.
To reduce redundant disk accesses, these methods often co-locate compressed vectors (e.g., PQ codes) with graph adjacency lists so that traversal signals and approximate distance estimates can be obtained from a single disk fetch.
However, since both traversal and scoring information are disk-resident, this design typically incurs more disk I/Os, which can degrade throughput and tail latency.


\noindent
\uline{\textbf{$\blacktriangleright$} Evaluation Preview.} 
In Section~\ref{sec:storage_evaluation}, we quantify the memory and disk footprints of both strategies and identify the memory threshold beyond which the major-in-disk design becomes infeasible and must fall back to an all-in-disk placement.

\subsection{Locality-Aware Disk Layout}

Careful layout of adjacency lists and raw vectors within disk blocks to improve data locality is critical for minimizing I/O cost. As shown in Figure~\ref{fig:layout}, layout decisions can be characterized at two levels: \emph{global} and \emph{local}. Global layout determines how adjacency lists and raw vectors are distributed across disk blocks, while the local layout specifies how these components are organized within each block.

\subsubsection{Global Layout}
Based on whether adjacency lists and raw vectors are co-located within the same disk blocks, existing global layouts can be classified into \emph{coupled} and \emph{decoupled} designs.

\begin{figure}[!t]
    \centering
    \includegraphics[width=\linewidth]{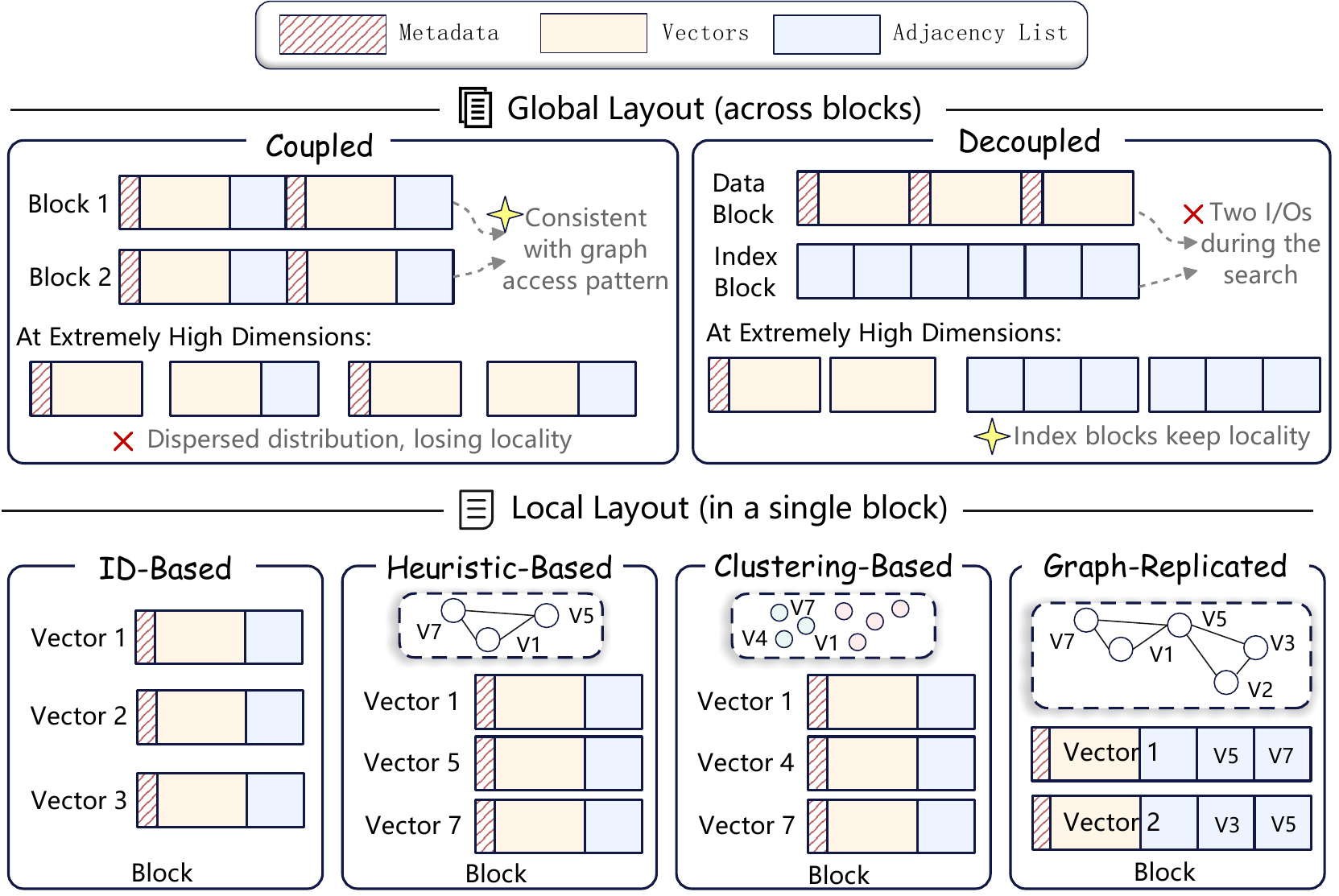}
   \vspace{-0.3in}
    \caption{Locality-aware disk layout strategies.}
   \vspace{-0.2in}
    \label{fig:layout}
\end{figure}

\noindent \textbf{Coupled Layout.} 
As illustrated in Figure~\ref{fig:layout}, the coupled layout colocates each raw vector (and its metadata) with its adjacency list within the same disk block, which is widely adopted by several methods, including DiskANN~\cite{DiskANN}, Starling~\cite{Starling}, and PageANN~\cite{PageANN}. 
This layout aligns well with graph-based ANN traversal, where result evaluation and neighbor expansion are performed together, making tight colocation natural for the access pattern. 
However, at very high dimensionality (e.g., $\geq 1000$), a disk block may hold only a single raw vector, leading to fragmented placement and degraded data locality, which ultimately limits I/O efficiency.

\noindent \textbf{Decoupled Layout.}
The decoupled layout stores adjacency lists and raw vectors separately in index blocks and data blocks, which is adopted by recent methods such as DGAI~\cite{DGAI}, TRIM~\cite{TRIM}, BAMG~\cite{BAMG}, and GaussDB-Vector~\cite{gaussdb-vector}. While decoupling adjacency lists from raw vectors risks doubling the I/O per traversal step, this penalty is mitigated by pruning strategies that bypass unnecessary data block accesses~\cite{DGAI,TRIM}. 
At high dimensionality, this layout achieves better locality: short adjacency lists allow many graph entries to be packed into a single index block, while pruning skips most data blocks, substantially reducing overall I/O cost.

\noindent
\uline{\textbf{$\blacktriangleright$} Evaluation Preview.} 
Our analysis indicates that data dimensionality plays a critical role in determining an appropriate global layout. 
In Section~\ref{sec:layout_evaluation}, we evaluate the impact of data dimensionality on global layout choices and derive practical guidelines for selecting an appropriate global layout under different dimensionalities.

\subsubsection{Local Layout} 
Local layout determines which vectors and their adjacency lists are colocated within the same disk block (under the coupled global layout), when a disk block can accommodate multiple data items as determined by the \emph{page size}. By improving data locality, an effective local layout enhances I/O effectiveness and reduces overall I/O cost.
As illustrated in Figure~\ref{fig:layout}, existing local layout strategies can be broadly classified into four categories: \emph{ID-based}~[\citealp{alayalaser}, \citealp{DiskANN,AiSAQ,LM-DiskANN,PipeANN}], \emph{heuristic-based}~\cite{Starling,margo}, \emph{clustering-based}~\cite{GoVector,PageANN,DGAI}, and \emph{graph-replicated}~\cite{Gorgeous}. 

\noindent 
\textbf{ID-Based.}
The ID-based layout organizes vector items within disk blocks based on their data identifiers, which is initially adopted by DiskANN~\cite{DiskANN}. This simple layout incurs no extra computational overhead, but it is largely oblivious to the graph search pattern. Vectors with consecutive IDs are not necessarily close in the graph, nor are they frequently accessed together during graph traversal. AlayaLaser~\cite{alayalaser} also follows this ID-based paradigm at the page
  level, while further improving intra-page efficiency by organizing neighbor quantization codes in a chunk-interleaved
  manner for better sequential scanning and SIMD utilization.

\noindent
\textbf{Heuristic-Based.} 
Heuristic-based layouts explicitly exploit graph topology and aim to colocate neighboring vectors within the same disk block to improve data locality. Prior work~\cite{Starling} shows that optimally serializing a graph into disk blocks while minimizing cross-block accesses is an NP-hard problem. As a result, they mainly rely on heuristics that incrementally place a vector into the disk block containing the largest number of its already-placed neighbors. Albeit without optimality guarantees, heuristic-based layouts can achieve better I/O effectiveness and reduced I/O cost in practice.

\noindent
\textbf{Clustering-Based.} This strategy groups vectors belonging to the same cluster into the same disk block, prioritizing vector similarity over graph topology.
This is motivated by the observation that long graph edges are often introduced to maintain connectivity or search diversity and thus do not reflect proximity in the original vector space~\cite{GoVector}.
By colocating similar vectors, clustering-based layouts can improve spatial locality and reduce I/O cost.

\begin{figure}[!t]
    \centering
    \includegraphics[width=\linewidth]{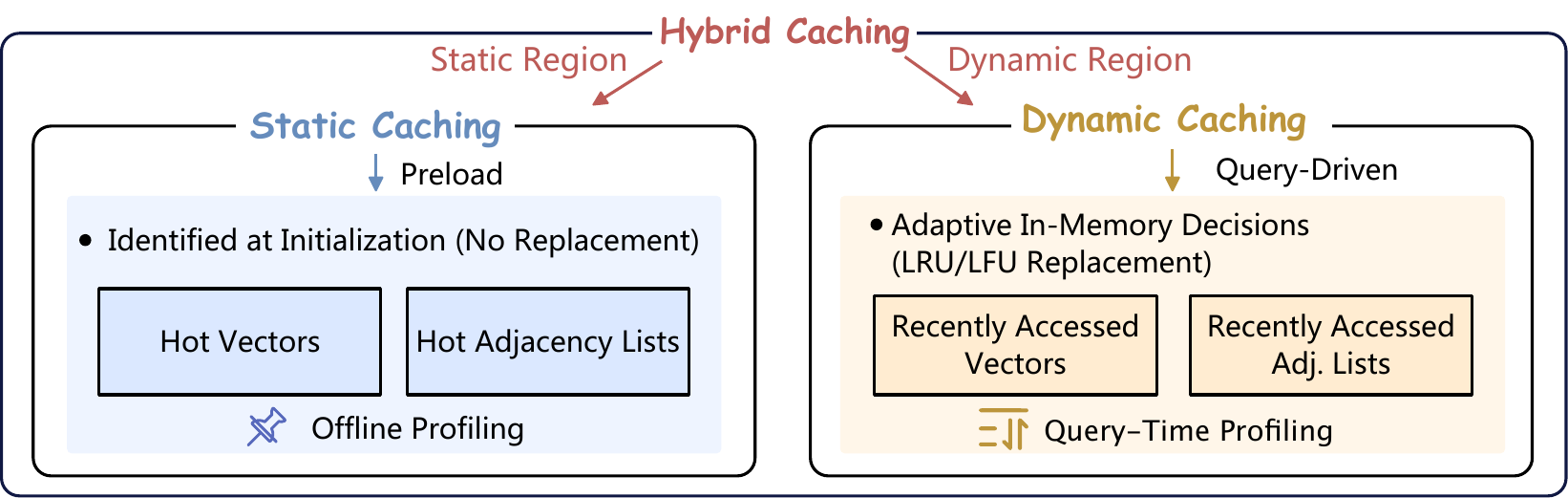}
   \vspace{-0.25in}
    \caption{Cache strategies.}
   \vspace{-0.25in}
    \label{fig:cache}
\end{figure}

\noindent
\textbf{Graph-Replicated.} 
Graph-replicated layout stores multiple adjacency lists together with a single raw vector in the disk block, motivated by the fact that adjacency lists are accessed far more frequently than raw vectors during graph traversal~\cite{Gorgeous}. As illustrated in Figure~\ref{fig:layout}, a graph-replicated block not only stores a raw vector (e.g., $v1$) together with its own adjacency list, but also includes the adjacency lists of its neighboring vectors (e.g., $v5$ and $v7$).
By duplicating neighbor information, this design reduces random I/O incurred by repeatedly fetching adjacency lists, at the cost of additional storage overhead due to replication.

\noindent
\uline{\textbf{$\blacktriangleright$} Evaluation Preview.} In Section~\ref{sec:layout_evaluation}, we evaluate the four classes of local layouts with a focus on their I/O effectiveness.
Specifically, we quantify how much of the raw vectors and adjacency lists read from disk are actually useful for graph traversal. In addition, since \emph{page\_size} is a key factor affecting data locality and query efficiency, we examine its impact separately in Section~\ref{sec:pagesize_evaluation}.

\subsection{Cache Management Strategy}
\label{subsec: cache_management}
Caches are commonly employed in disk-resident ANN search to keep frequently accessed or potentially useful data in memory, thereby reducing disk I/O. As shown in Figure~\ref{fig:cache}, we classify existing cache management strategies into three categories:
static~[\citealp{alayalaser}, \citealp{DiskANN, Gorgeous, Starling, PageANN}], dynamic~\cite{XN-Graph}, and hybrid caching strategies~\cite{GoVector, DGAI}.

\noindent \textbf{Static Caching.} 
Static caching pre-identifies frequently accessed \emph{hot} data and retains them in memory. Such data typically include hub vectors important for graph navigation or answer vectors for popular or recurring queries. Caching them can significantly reduce disk I/O, as most vectors are rarely accessed~\cite{turbopuffer,ZillizCloud}. However, static caching is sensitive to workload skew and may incur high latency for cold or unseen queries that deviate from the cached hot set.

\noindent \textbf{Dynamic Caching.} 
Dynamic caching adaptively retains raw vectors and adjacency lists in memory based on runtime access patterns. Using online replacement policies (e.g., LRU or LFU), it captures temporal locality and adapts to evolving workloads, reducing latency for changing or unseen queries. However, cache maintenance adds computational overhead and becomes less effective under highly random access patterns.

\noindent \textbf{Hybrid Caching.} 
Hybrid caching combines static and dynamic caching by partitioning memory into a static region for long-term hot items and a dynamic region managed by online replacement policies to capture transient access patterns.
This design balances throughput and latency across diverse query workloads, but its effectiveness depends on proper memory partitioning and tuning.

\begin{figure}[!t]
    \centering
    \includegraphics[width=\linewidth]{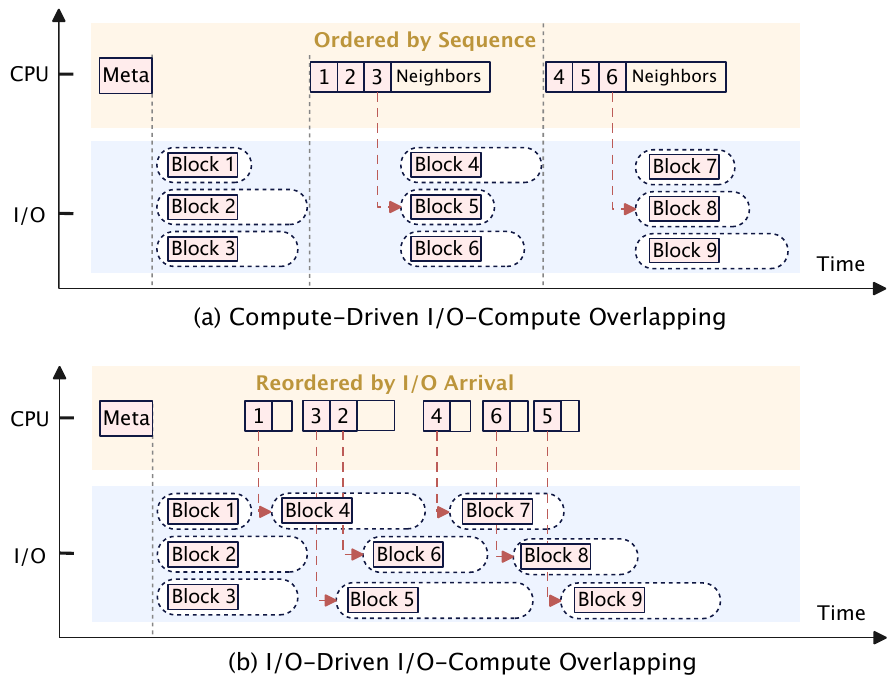}
   \vspace{-0.3in}
    \caption{Compute-driven vs. I/O-driven overlapping.}
   \vspace{-0.2in}
    \label{fig:asynchronous}
\end{figure}

\noindent
\uline{\textbf{$\blacktriangleright$} Evaluation Preview.} In Section~\ref{sec:cache_evaluation}, we conduct a comparison of the three cache management strategies, examining which design achieves higher cache utilization.

\subsection{Query Execution}
Query execution in disk-resident graph-based ANN search largely follows a standard best-first graph traversal process, as described in Section~\ref{sec:preliminary}. In this workflow, memory-resident compressed vectors are typically used for distance estimation, while disk-resident adjacency lists and full-precision vectors are accessed for candidate expansion and result evaluation. Depending on the adopted storage, layout, and caching strategies, existing methods introduce targeted modifications to this baseline procedure. For example, \emph{all-in-disk} designs perform distance estimation using disk-resident compressed vectors~\cite{AiSAQ,LM-DiskANN,PageANN}, \emph{decoupled layouts} separate navigation and refinement with additional pruning to reduce data accesses~\cite{DGAI,TRIM}, and \emph{dynamic caching} adaptively determines which vectors and adjacency lists to retain in memory at runtime~\cite{XN-Graph,GoVector,DGAI}.

To mitigate high latency of disk accesses, modern systems widely adopt asynchronous execution to overlap I/O operations with CPU computation, reducing I/O stall time. Depending on whether subsequent I/O requests are determined by intermediate computation results or issued speculatively and filtered by later computation, existing I/O–compute overlap strategies can be broadly classified into two categories: \emph{compute-driven}~\cite{Starling,Gorgeous,PageANN}and \emph{I/O-driven}~\cite{alayalaser,PipeANN}.
\noindent\textbf{Compute-Driven I/O--Compute Overlapping.}
As shown in Figure~\ref{fig:asynchronous}(a), compute-driven overlapping follows a CPU-determined execution order. After reading metadata, the CPU processes candidate vectors sequentially within a limited-width beam and issues I/O requests only after computation identifies the required data blocks. Therefore, disk accesses are tightly coupled with computation results and fetched on demand, matching graph traversal patterns while avoiding unnecessary reads. However, since each I/O request depends on prior computation, the degree of overlapping is limited, potentially underutilizing available I/O bandwidth.

\noindent\textbf{I/O-Driven I/O–Compute Overlapping.} 
Figure~\ref{fig:asynchronous}(b) illustrates an I/O-driven strategy, where disk requests are issued speculatively and decoupled from the computation order. After metadata access, multiple I/O requests for candidate blocks, determined by the beam width, are submitted proactively. As blocks return asynchronously, the CPU processes available data immediately, reordering computation by I/O arrival rather than traversal sequence. This maximizes overlap and bandwidth utilization but may cause redundant disk reads and extra memory pressure due to speculative prefetching.

\noindent
\uline{\textbf{$\blacktriangleright$} Evaluation Preview.} 
In Section~\ref{sec:overall_performance}, we evaluate the end-to-end performance of representative methods across different storage, layout, and caching strategies, and analyze their sensitivity to \emph{beam\_width}. Section~\ref{sec:query_evaluation} further examines CPU and I/O efficiency using distance-computation time and the number of disk I/Os.

\subsection{Update Mechanism}
Under dynamic workloads, disk-resident graph-based ANN methods support continuous insertions and deletions while targeting three goals: (1) preserving graph connectivity, (2) maintaining query efficiency, and (3) minimizing update overhead.
Based on when updates are materialized, existing methods can be classified into \emph{in-place}~\cite{LM-DiskANN,DGAI,pgvector,gaussdb-vector,ODINANN} and \emph{out-of-place}~\cite{Milvus,LSM-VEC,Aster,LEANN, FreshDiskANN} updates.

\noindent
\textbf{In-Place Update.}
In-place designs apply modifications directly to the disk-resident graph as updates arrive. A typical in-place mechanism is to edit on-disk adjacency immediately and perform local graph topology maintenance (e.g., linking, pruning, and reconnection) so that the graph remains searchable right after each update~\cite{LM-DiskANN,DGAI}. Another in-place variant still writes updates to disk on the critical path, but avoids doing \emph{all} cleanup work synchronously: deletions may be recorded in-place while space reclamation and structural tidying are deferred to maintenance, without introducing a separate memory-resident delta that must later be merged into the main index~\cite{pgvector,gaussdb-vector,ODINANN}. This strategy provides strong freshness and avoids a multi-component index state, but its cost is sensitive to random I/O and write interference, as a single update may touch multiple pages and contend with foreground search.

\noindent
\textbf{Out-of-Place Update.}
Out-of-place designs first accumulate modifications in memory and then write them to disk in a batched manner. A representative implementation is an LSM-style consolidation pipeline, where updates are absorbed into memory and periodically merged/compacted into a disk-resident graph index~\cite{Milvus,LSM-VEC,Aster,FreshDiskANN}. Another implementation maintains a memory-resident overlay that is immediately visible to queries and asynchronously integrated into the base graph index, so that updates become searchable without immediately rewriting the disk graph~\cite{LEANN}. This strategy improves write throughput by amortizing many small changes into batched disk writes and better isolating reads from writes, but it introduces a multi-component index state and costly background consolidation.

\noindent
\uline{\textbf{$\blacktriangleright$} Evaluation Preview.}
In Section~\ref{sec:update_evaluation}, we compare in-place and out-of-place updates in terms of update latency and analyze how query performance evolves as updates accumulate.

\begin{figure*}[t]
    \centering
    \includegraphics[width=0.9\linewidth]{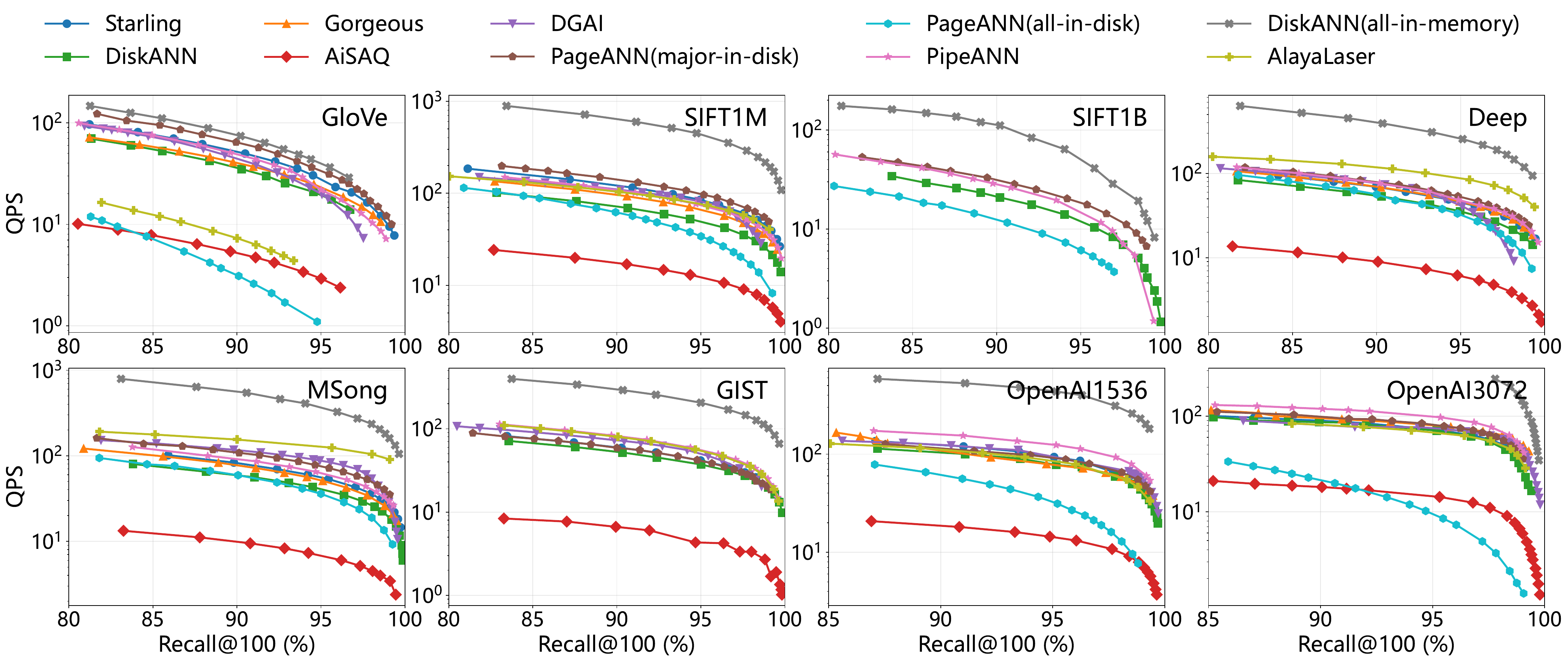}
     \vspace{-0.15in}
    \caption{Trade-offs between QPS and recall.}
     \vspace{-0.1in}
    \label{Fig:overall_qps_recall}
\end{figure*}

\section{Experiments}
\label{sec:experiments}
\subsection{Experimental Setup}
\label{sec:exp_setup}

\noindent
\textbf{Datasets.} 
As summarized in Table~\ref{tab:datasets}, we evaluate our methods on eight benchmark vector datasets~\cite{texmex-datasets,gqr-datasets,qdrant_dbpedia_openai} spanning diverse scales and dimensionalities.
End-to-end QPS and recall are evaluated on all datasets, while other experiments are mainly reported on SIFT1M, Deep, and GIST. These datasets cover low-, medium-, and high-dimensional spaces, dictating that a standard 4KB disk page can store  5, 3, and 1 nodes, respectively.
This provides a robust basis to evaluate the performance characteristics of different design choices.

\begin{table}[h]
\centering
\vspace{-0.1in}
\caption{Dataset statistics and parameters. $R$: maximum degree; Samp.: graph sampling rate; Cache: cache budget (fraction of dataset size); Page: page size; PQ $m$: subspace count.} 
\vspace{-0.15in}
\label{tab:datasets}
\footnotesize
\renewcommand{\arraystretch}{1.05}
\setlength{\tabcolsep}{2pt}

\begin{tabular*}{\columnwidth}{@{\extracolsep{\fill}}lccccccccc}
\toprule
& \multicolumn{4}{c}{Dataset Statistics} 
& \multicolumn{5}{c}{Parameter Settings} \\
\cmidrule(lr){2-5} \cmidrule(lr){6-10}
Dataset 
& Dim & Type & \#Vectors & \#Queries 
& $R$ & Samp. & Cache & Page & PQ $m$ \\
\midrule
GloVe    & 100   & float & 1,183,514     & 1,000  
         & 48 & 0.5\% & 1\% & 4KB  & 25 \\

SIFT1M   & 128   & float & 1,000,000     & 10,000 
         & 48 & 0.5\% & 1\% & 4KB  & 16 \\

SIFT1B   & 128   & uint8 & 1,000,000,000 & 10,000 
         & 48 & 0.5\% & 1\% & 4KB  & 16 \\

Deep     & 256   & float & 1,000,000     & 1,000  
         & 48 & 0.5\% & 1\% & 4KB  & 32 \\

MSong    & 420   & float & 992,272       & 1,000  
         & 48 & 0.5\% & 1\% & 4KB  & 64 \\

GIST     & 960   & float & 1,000,000     & 1,000  
         & 48 & 0.5\% & 1\% & 4KB  & 120 \\
         
OpenAI1536   & 1,536 & float & 999,000   & 1,000  
         & 48 & 0.5\% & 1\% & 8KB &  192\\
         
OpenAI3072   & 3,072 & float & 1,000,000 & 1,000  
         & 48 & 0.5\% & 1\% & 16KB & 384 \\
\bottomrule
\end{tabular*}

\vspace{-0.1in}
\end{table}


\noindent
\textbf{Compared Methods.}
As shown in Table~\ref{tab:compared_methods}, we evaluate 8 representative open-source methods covering diverse storage, layout, caching, query execution, and update designs: DiskANN~\cite{DiskANN} and its update-enhanced variant FreshDiskANN~\cite{FreshDiskANN}, AiSAQ~\cite{AiSAQ}, Starling~\cite{Starling}, Gorgeous~\cite{Gorgeous}, PageANN~\cite{PageANN}, PipeANN~\cite{PipeANN} and its update-enhanced variant OdinANN~\cite{ODINANN}, DGAI~\cite{DGAI} and AlayaLaser~\cite{alayalaser}. AlayaLaser differs from the other methods by adopting RaBitQ~\cite{gao2024rabitq} for vector quantization and padding each node's neighbor list to a multiple of 32. Following its original setting, we use $R=64$ instead of $48$. Its page size is automatically determined by its layout and may differ from the defaults in Table~\ref{tab:datasets}.

\noindent
\textbf{Implementations.} 
We integrate and extend the available upstream open-source implementations ~\cite{alayalitegithub,pipeanngithub,dgaigithub,aisaq-diskanngithub,FreshDISKANNgithub,gorgeousgithub,pageanngithub} of the compared methods in C++ (GCC -O3) with unified evaluation workflows and instrumentation.
Experiments are conducted on a Linux server equipped with dual Intel Xeon Platinum 8457C processors (48 cores/96 threads each) and 512~GB RAM. Such server-class CPUs are well suited to large-scale vector search workloads involving compute-intensive operations such as high-dimensional distance computation and index construction. All static search experiments use a single CPU thread, while multiple threads are used only for index construction, dynamic workloads, and dedicated multi-thread scalability experiments. 
Following prior disk-resident ANN studies~\cite{Starling,PipeANN,FreshDiskANN,PageANN}, we use a single Samsung 970 EVO 2~TB NVMe SSD (\seqsplit{MZ-V7E2T0BW}), ensuring that the measured performance primarily reflects the I/O efficiency of each ANN method without additional bandwidth or device-level parallelism from multiple SSDs. The SSD uses a PCIe 3.0 $\times$4 interface with a Linux queue depth of 256, providing a sequential-read bandwidth of up to 3,500~MB/s and 4~KB random-read performance of 15K IOPS at QD1 (1 thread) and up to 500K IOPS at QD32 (4 threads). 
\subsection{End-to-End Evaluation}
\label{sec:overall_performance}

\subsubsection{Trade-offs between QPS and Recall.}
\label{sec:QPSRecall}

Figure~\ref{Fig:overall_qps_recall} compares end-to-end query performance across eight datasets. 
Note that several algorithms are excluded from the evaluation on SIFT1B due to index construction failures, as detailed in Section~\ref{sec:index_evaluation}. Additionally, Gorgeous and PageANN (all-in-disk) could not be built on GIST, as discussed in Section~\ref{sec:pagesize_evaluation}.
Overall, PageANN (major-in-disk) excels in low dimensions, while PipeANN dominates high-dimensional GIST and OpenAI. For example, at 0.95 recall, PageANN (major-in-disk) outperforms PipeANN by 42.8\% on GloVe, 25.7\% on SIFT1M, 22.3\% on SIFT1B. In high dimensions, PipeANN surpasses AlayaLaser by 3.1\% on GIST, 57.0\% on OpenAI1536, and 36.8\% on OpenAI3072. PageANN's advantage in low dimensions stems from a lightweight in-memory hash table for efficient entry selection and a block-aligned graph index that maximizes useful data per I/O. As dimensionality grows, PipeANN's I/O–compute overlapping reduces latency from multiple block accesses. 
On MSong and Deep, AlayaLaser becomes the best performing method, exceeding 41.4\% and 57.1\%, respectively. This is mainly because AlayaLaser's overall error (from both PCA and RaBitQ) is smaller than PQ-induced errors of other methods on these datasets, enabling faster and more accurate graph search.


Across benchmarks, the in-memory baseline consistently outperforms disk-based methods by 1.4$\times$--4.5$\times$. Major-in-disk designs generally outperform all-in-disk variants by trading memory for fewer disk accesses.  On average, this trade-off is more effective on GloVe and GIST, where 2.0$\times$--3.5$\times$ more memory yields 1.5$\times$--7.2$\times$ higher QPS, than on OpenAI3072, where 4.9$\times$ more memory provides only 2.4$\times$ speedups.



\vspace{-0.1in}
\begin{tcolorbox}[
  colback=black!5,
  colframe=white,
  boxrule=0pt,
  arc=2pt,
  left=6pt,right=6pt,top=4pt,bottom=4pt
]
\label{findings:QPSRecall}
\textbf{Findings}: 
(1) PageANN (major-in-disk) outperforms in low dimensions, whereas PipeANN dominates in high dimensions, underscoring the need for \emph{dimension-aware design choices}. In high-dimensional settings, poor block locality makes multi-block accesses inevitable, favoring I/O-driven overlapping to 
hide read latency. 
In contrast, at lower dimensions, stronger block locality makes graph topology and layout more critical for reducing block accesses.
(2) \emph{Major-in-disk} designs generally outperform \emph{all-in-disk} variants, highlighting the benefits of keeping compressed vectors in memory.
(3) On some datasets, the combination of PCA and RaBitQ incurs less accuracy loss than PQ, enabling AlayaLaser to achieve better performance.
\end{tcolorbox} 

\subsubsection{Index Construction Time and Peak Memory Overhead.}
\label{sec:index_evaluation}

\begin{figure}[H]
    \centering
    \includegraphics[width=\linewidth]{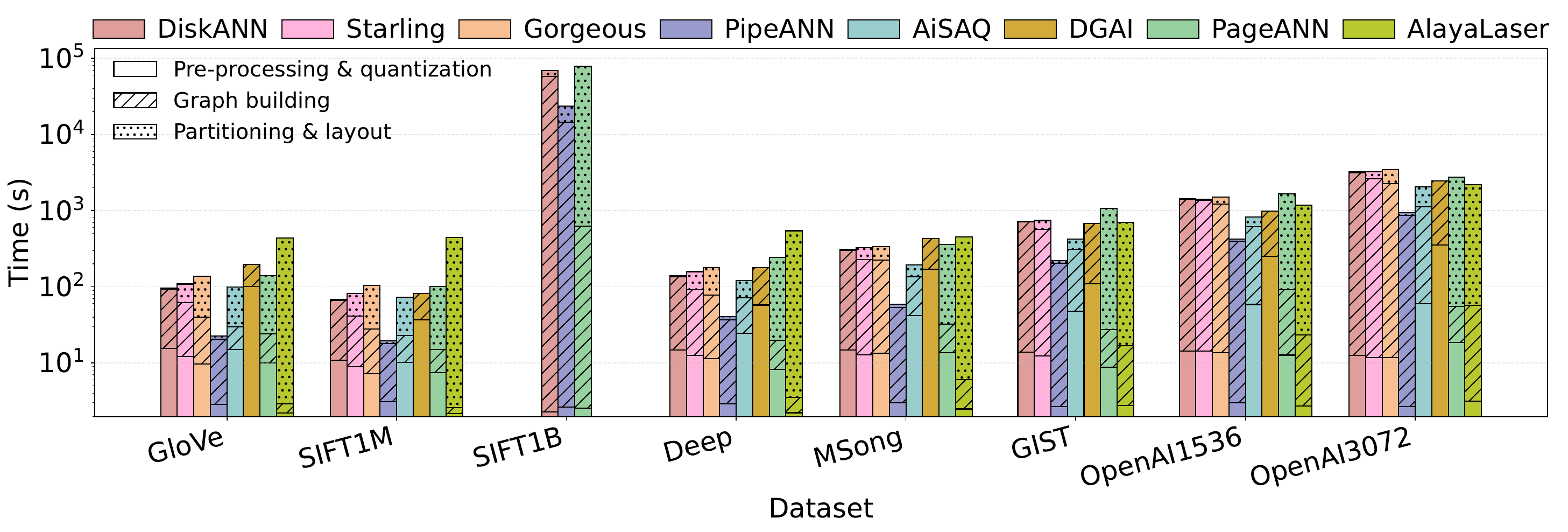}
     \setlength{\abovecaptionskip}{-6pt}
     \vspace{-0.1in}
     \caption{Index construction time.}
     \vspace{-0.2in}
    \label{fig:index_construction_time}
\end{figure}

Figure~\ref{fig:index_construction_time} reports the index construction time of eight algorithms across eight datasets, broken down into three stages: pre-processing \& quantization, graph building, and partitioning \& layout.
On SIFT1B, several methods fail to complete: DGAI exceeds the memory budget during graph construction; Starling incurs excessive memory usage during partitioning and layout; Gorgeous and AlayaLaser exceed SSD capacity due to their specialized layout; and AiSAQ suffers prohibitively long serial disk writing.
For most methods, construction time is dominated by graph building and pre-processing \& quantization, whereas PageANN and AlayaLaser incur substantial partitioning \& layout overhead. PageANN spends 45.6\% of its total time on graph partitioning on SIFT1B, and AlayaLaser spends 51.9\% of its total time on disk layout on OpenAI3072.
PipeANN achieves the shortest construction time across all datasets and is particularly efficient on OpenAI3072, outperforming AlayaLaser, AiSAQ and PageANN by 2.6$\times$--2.9$\times$, and DiskANN, Starling and Gorgeous by 3.5$\times$--3.7$\times$. Its advantage comes from efficient Elkan-based quantization~\cite{elkan}, single-pass graph construction with lightweight cleanup, and an ID-based layout that avoids graph partitioning.

\vspace{-0.1in}

\begin{figure}[H]
    \centering
    \includegraphics[width=\linewidth]{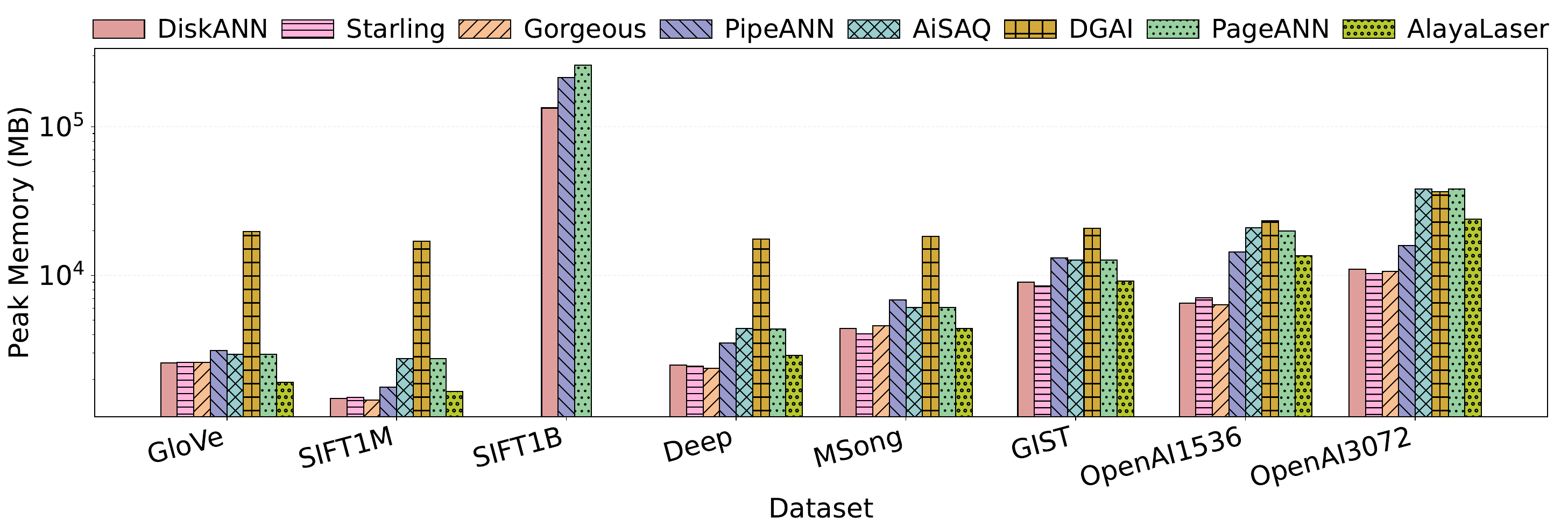}
     \setlength{\abovecaptionskip}{-6pt}
     \vspace{-0.1in}
     \caption{Peak memory during index construction.}
    \vspace{-0.15in}
    \label{fig:index_construction_memory}
\end{figure}

Figure~\ref{fig:index_construction_memory} reports peak memory during construction. Notably, DGAI shows the highest peak memory on most datasets, mainly because its HPQ building maintains temporary buffers whose size scales primarily with the number of data points.

\vspace{-0.15in}



\vspace{-0.05in}
\begin{tcolorbox}[
  colback=black!5,
  colframe=white,
  boxrule=0pt,
  arc=2pt,
  left=6pt,right=6pt,top=4pt,bottom=4pt
]
\textbf{Findings:} 
(1) For most methods, index construction time is mainly consumed by graph building and pre-processing \& quantization. Notable exceptions are PageANN and AlayaLaser, where partitioning \& layout cost accounts for most of the time.

(2) PipeANN achieves the highest construction efficiency by optimizing all three stages: it accelerates PQ training using the Elkan algorithm~\cite{elkan}, streamlines graph building via a single-pass global linking, and eliminates layout overhead via an ID-based layout strategy.

\end{tcolorbox} 
\vspace{-0.1in}

\subsubsection{Impact of Page Size}
\label{sec:pagesize_evaluation}
\begin{figure}[h]
    \includegraphics[width=0.85\linewidth]{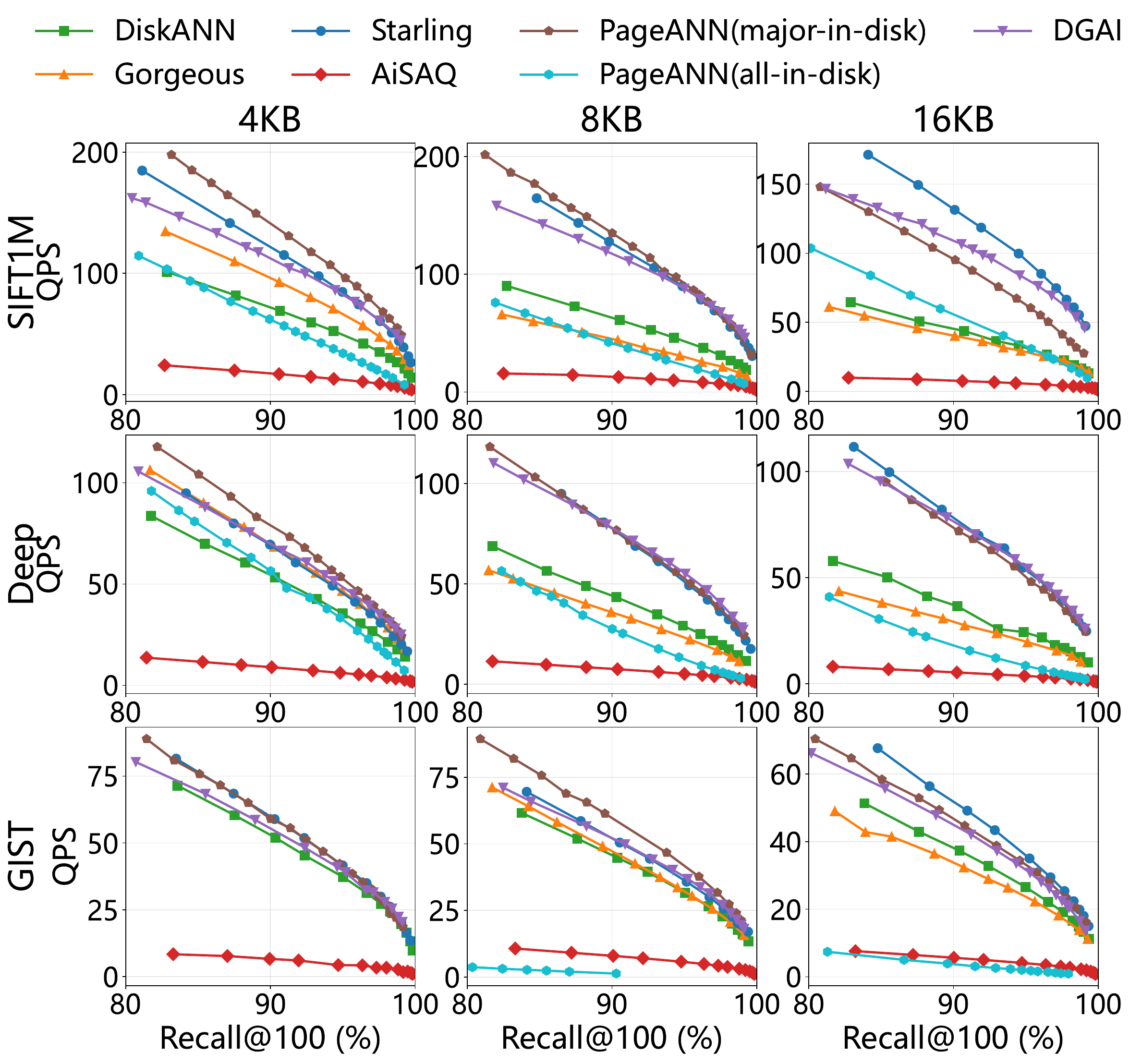}
   \vspace{-0.1in}
    \caption{QPS vs. recall for different page sizes.}
    \vspace{-0.15in}
    \label{Fig:page_size}
\end{figure}

Figure~\ref{Fig:page_size} shows the impact of page size on query efficiency. 
For most methods, QPS drops as page size increases from 4 KB to 16 KB. 
On SIFT1M, PageANN (major-in-disk) achieves the best performance at 4 KB and 8 KB page sizes, outperforming the runner-up by 11.2\%–25.8\% and 5\%–9\% at equal recall, respectively. However, at 16 KB, Starling becomes the top-performing method, exceeding the runner-up by 2.8\%–28\%. 
This trend stems from their different layout strategies. At small page sizes, PageANN benefits from its clustering-based layout with node contraction and neighbor pruning, which improves page utilization and access locality. However, as page size increases, its capped neighbor limit reduces the average out-degree and weakens graph connectivity. In contrast, Starling employs a heuristic-based layout without a strict neighbor cap, preserving richer connectivity and thus achieving better performance at larger page sizes. Detailed comparisons are provided in Section~\ref{sec:layout_evaluation}.

Note that \emph{not all methods can be built across all page sizes}. For example, on GIST with 4 KB pages, Gorgeous fails to construct the graph index with $R=48$, while PageANN (all-in-disk) becomes highly inefficient because the limited page capacity restricts each node to only two neighbors. Additionally, PipeANN and AlayaLaser automatically determine their optimal page sizes, avoiding manual tuning and are therefore excluded from this experiment. For the remaining methods, smaller page sizes generally yield higher efficiency across dimensionalities. Larger pages incur longer I/O latency and provide more data per access. Without an effective layout strategy to exploit this extra data, most of it becomes redundant, introducing extra I/Os. Section~\ref{sec:layout_evaluation} shows that all layout strategies exhibit low I/O effectiveness,
limiting the benefits of larger pages.

\vspace{-0.05in}
\begin{tcolorbox}[
  colback=black!5,
  colframe=white,
  boxrule=0pt,
  arc=2pt,
  left=6pt,right=6pt,top=4pt,bottom=4pt
]
 
\textbf{Findings:} 
(1) PageANN (major-in-disk) performs best under small page sizes (4KB/8KB), while Starling dominates at 16KB. This suggests that with small page sizes, selecting high-quality neighbors is critical, whereas with larger page sizes, retaining more neighbors becomes advantageous.
(2) Some methods become infeasible under small page sizes since a single page cannot accommodate their required data structures. (3) When feasible, \emph{smaller page sizes} generally yield higher efficiency across different dimensionalities.
\end{tcolorbox} 

\vspace{-0.2in}
\begin{figure}[H]
    \includegraphics[width=\linewidth]{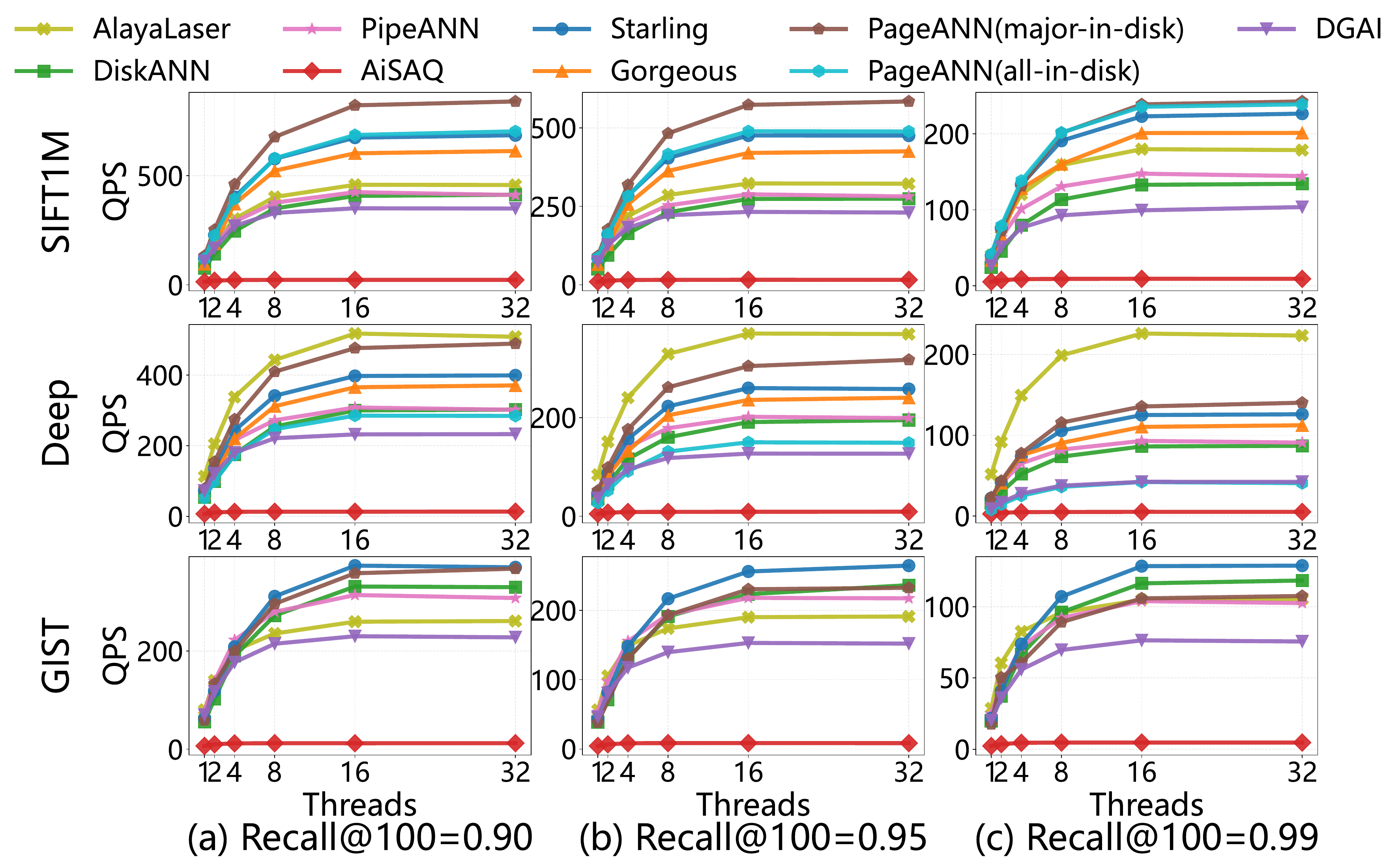}
    \vspace{-0.05in}
   \setlength{\abovecaptionskip}{-6pt}
    \caption{QPS vs. recall for different thread count.}
    \vspace{-0.1in}
    \label{Fig:thread_count}
\end{figure}

\subsubsection{Impact of thread count}
\label{subsubsec:thread_count}
Figure~\ref{Fig:thread_count} shows the impact of varying the number of search threads on query efficiency. On SIFT1M  at 0.95 recall, scaling from 1 to 8 threads yields a 4.73$\times$--5.27$\times$ QPS speedup for methods like Starling and PageANN (major-in-disk). However, further increasing to 16 threads provides only a marginal 17.8\%--18.8\% gain. Despite doubling the thread count from 16 to 32, the QPS improvement for most algorithms falls below 1\%, while PipeANN even experiences a 2.36\% performance degradation. This bottleneck occurs because while higher concurrency initially improves throughput by parallelizing search requests, the aggregate I/O demands eventually saturate the underlying SSD's bandwidth.


\vspace{-0.05in}
\begin{tcolorbox}[
  colback=black!5,
  colframe=white,
  boxrule=0pt,
  arc=2pt,
  left=6pt,right=6pt,top=4pt,bottom=4pt
]
\textbf{Findings:} 
Increasing the thread count improves search throughput, but the gains eventually plateau due to SSD I/O saturation, demonstrating that high-concurrency ANNS workloads are fundamentally I/O-bound.
\end{tcolorbox} 
\vspace{-0.1in}

\subsubsection{Impact of static Beam Width}
\label{sec:Beamwidth_evaluation}
\begin{figure}[H]
    \centering
    \includegraphics[width=\linewidth]{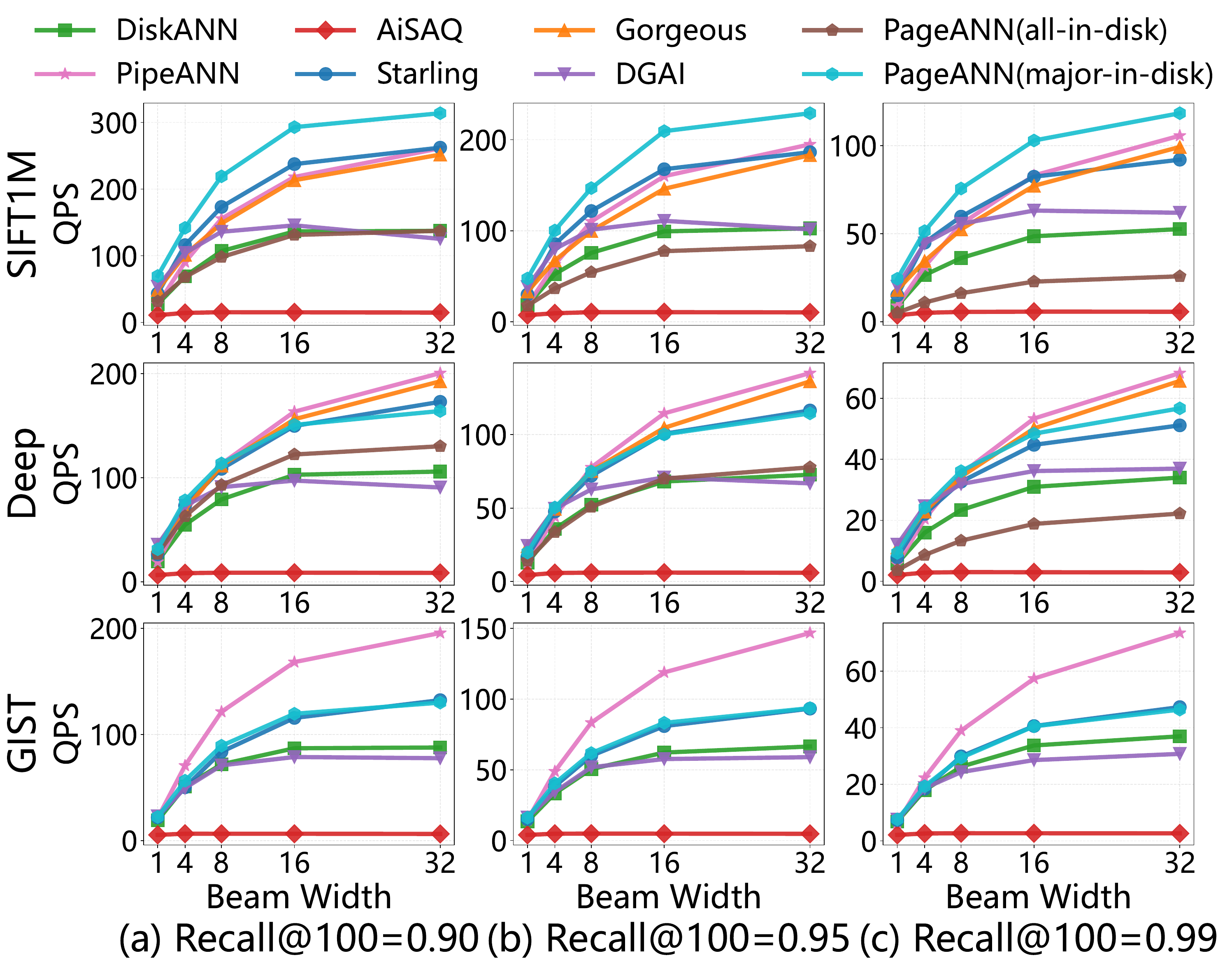}
    \vspace{-0.3in}
    \caption{QPS vs. recall for different static beam widths.}
    \vspace{-0.1in}
    \label{Fig:beam_width}
\end{figure}

Figure~\ref{Fig:beam_width} shows the impact of varying the static beam width ($W$) on query efficiency across datasets and recall targets. Overall, QPS increases with larger beam widths, with the largest gains occurring between $W=1$ and $W=8$.
For example, on SIFT1M at 0.95 recall, increasing $W$ from 1 to 4 improves QPS by approximately $2.8\times$–$3.4\times$ for methods such as DiskANN and PipeANN, while increasing $W$ from 4 to 8 still yields $45\%$–$75\%$ gains. However, the benefit diminishes significantly beyond $W=16$, where QPS improvements for most methods fall below 15\% or nearly plateau. 
This reveals a fundamental trade-off: larger beam widths initially improve throughput through better I/O batching, but the gains are eventually limited by SSD I/O concurrency and the overhead of maintaining larger candidate pools. Notably, the saturation point also depends on the underlying storage hardware. SSDs with higher I/O concurrency can sustain the benefits of increasing $W$ over a wider range, thereby shifting the saturation threshold toward larger beam widths.

\vspace{-0.1in}

Beyond static settings, both PipeANN\cite{PipeANN} and AlayaLaser\cite{alayalaser} utilize adaptive beam widths to optimize performance. PipeANN dynamically expands its width based on the candidate pool state to reduce early I/O waste, while AlayaLaser grows it exponentially with hop counts to focus exploration near the target. Both approaches set their upper limit for dynamic beam width around 16 or 32, which aligns with our findings in the exploration.

\vspace{-0.05in}
\begin{tcolorbox}[
  colback=black!5,
  colframe=white,
  boxrule=0pt,
  arc=2pt,
  left=6pt,right=6pt,top=4pt,bottom=4pt
]
\textbf{Findings:} Increasing the beam width improves query performance, but the marginal gains progressively diminish, plateauing at around 16 in our settings.
\end{tcolorbox} 
\vspace{-0.15in}

\definecolor{myblue}{RGB}{0,102,204}
\definecolor{myred}{RGB}{220,50,47}
\begin{table*}[t]
\centering
\caption{Disk (GB) and memory (MB) usage across datasets (min/max highlighted \emph{per dataset} across all methods).}
\vspace{-0.15in}
\label{tab:type_dataset_disk_mem}
\small
\renewcommand{\arraystretch}{1.15}
\setlength{\tabcolsep}{4.2pt}
\resizebox{0.9\textwidth}{!}{
\begin{tabular}{l|l|cc|cc|cc|cc|cc|cc|cc|cc}
\toprule
\multirow{2}{*}{\textbf{Type}} & \multirow{2}{*}{\textbf{Method}} &
\multicolumn{2}{c|}{\textbf{GloVe}} &
\multicolumn{2}{c|}{\textbf{SIFT-1M}} &
\multicolumn{2}{c|}{\textbf{SIFT-1B}} &
\multicolumn{2}{c|}{\textbf{Deep}} &
\multicolumn{2}{c|}{\textbf{MSong}} &
\multicolumn{2}{c|}{\textbf{GIST}} &
\multicolumn{2}{c|}{\textbf{OpenAI1536}} &
\multicolumn{2}{c}{\textbf{OpenAI3072}} \\
\cmidrule(lr){3-4}\cmidrule(lr){5-6}\cmidrule(lr){7-8}\cmidrule(lr){9-10}
\cmidrule(lr){11-12}\cmidrule(lr){13-14}\cmidrule(lr){15-16}\cmidrule(lr){17-18}
& & \textbf{Disk} & \textbf{Mem} & \textbf{Disk} & \textbf{Mem} & \textbf{Disk} & \textbf{Mem}
  & \textbf{Disk} & \textbf{Mem} & \textbf{Disk} & \textbf{Mem} & \textbf{Disk} & \textbf{Mem}
  & \textbf{Disk} & \textbf{Mem} & \textbf{Disk} & \textbf{Mem} \\
\midrule

\multirow{6}{*}{\textbf{Major-in-Disk}} &
DiskANN
& 0.790 & 160
& 0.811 & 108
& 349 & \cellcolor{myred!20}26465
& 1.368 & 175
& 2.052 & 279
& 4.174 & 574
& 8.194& 760
& 16.399 & 933 \\
& Starling
& 0.790 & 123
& 0.811 & 120
& -- & --
& 1.368 & 196
& 2.052 & 306
& 4.174 & 628
& 8.194& 898
& 16.399 & 1039 \\
& Gorgeous
& \cellcolor{myred!20}5.493 & 138
& \cellcolor{myred!20}4.768 & 109
& -- & --
& \cellcolor{myred!20}5.277 & 198
& 5.878 & 193
& -- & --
& \cellcolor{myred!20}13.530& 754
& \cellcolor{myred!20}30.708 & 818 \\
& PipeANN
& 0.752 & 153
& 0.763 & 125
& 318 &  23008
& 1.272 & 155
& \cellcolor{myblue!25}1.896 & 194
& \cellcolor{myblue!25}3.815 & 274
& \cellcolor{myblue!25}7.622& 344
& \cellcolor{myblue!25}15.259 & 749.5 \\
& DGAI
& 0.864 & \cellcolor{myred!20}556
& 0.862 & \cellcolor{myred!20}454
& -- & --
& 1.545 & \cellcolor{myred!20}586
& 2.650 & \cellcolor{myred!20}829
& 4.001 & \cellcolor{myred!20}958
& 7.627& \cellcolor{myred!20} 1383
& 16.254 & \cellcolor{myred!20}2014 \\
& PageANN$^{\dagger}$
& 0.613 & 102
& 0.605 & 61.5
& \cellcolor{myblue!25}212 & 19384
& 1.367 & 88.5
& 2.053 & 100.5
& 4.186 & 169.5
& 7.626& 217.5
& 16.529 & 456 \\

\midrule

\multirow{2}{*}{\textbf{All-in-Disk}} &
AiSAQ
& 1.570 & \cellcolor{myblue!25}33
& 1.335 & \cellcolor{myblue!25}30
& -- & --
& 2.033 & 31.5
& 2.118 & \cellcolor{myblue!25}31.5
& 4.286 & \cellcolor{myblue!25}36
& 8.372& 44.5
& 16.408 & 48 \\
& PageANN$^{\ddagger}$
& \cellcolor{myblue!25}0.514 & 43.5
& \cellcolor{myblue!25}0.558 & \cellcolor{myblue!25}30
& \cellcolor{myred!20}424 & \cellcolor{myblue!25}30.6
& \cellcolor{myblue!25}1.270 & \cellcolor{myblue!25}30
& 1.900 & 34.5
& -- & --
& \cellcolor{myblue!25}7.622& \cellcolor{myblue!25}36
& 15.260 & \cellcolor{myblue!25}45 \\
& AlayaLaser
& 2.257 
& 126
& 1.907
& 113
& -- 
& --
& 3.815
& 120
& \cellcolor{myred!20}7.585 
& 140
& \cellcolor{myred!20}11.444
& 196
& 11.433 
& 236
& 22.888 
& 396 \\
\bottomrule
\end{tabular}
}
\vspace{2.5pt}

\footnotesize\raggedright
\hspace*{0.04\textwidth}
\textbf{Notes:} \cellcolor{myblue!25}{blue} = minimum, \cellcolor{myred!20}{red} = maximum \textbf{within each dataset and metric} (Disk or Mem) across all methods; ties are highlighted; missing entries (\texttt{--}) are excluded. \par
\hspace*{0.04\textwidth}
$^{\dagger}$ PageANN major-in-disk version; $^{\ddagger}$ PageANN all-in-disk (low-memory) version.
\vspace{-0.1in}
\end{table*}

\subsection{Fine-Grained Evaluation }
\label{sec:fine_performance}
\subsubsection{Storage Strategy Evaluation}
\label{sec:storage_evaluation}
Table~\ref{tab:type_dataset_disk_mem} reports the disk (GB) and memory (MB) footprints of disk-resident graph-based ANN methods across eight datasets. Major-in-disk strategies require substantially more memory, whereas all-in-disk methods maintain a nearly constant footprint of only $25.5$--$48$ MB regardless of dataset scale except AlayaLaser. AlayaLaser incurs relatively high memory overhead because it maintains multiple in-memory components, including entry points, cached nodes, and auxiliary data structures. DGAI incurs the highest memory overhead (up to $2014.1$ MB on OpenAI3072) because HPQ reranking requires multiple in-memory auxiliary structures, (e.g.,codebooks). Excluding DGAI, major-in-disk methods consume $3.89\%$--$35.44\%$ of the raw dataset size in memory, with the highest overheads typically observed on low-dimensional datasets. Starling exhibits the largest memory footprint on high-dimensional datasets, reaching $628$ MB on GIST and $1039$ MB on OpenAI3072. This large memory overhead mainly stems from maintaining massive auxiliary structures in memory, such as PQ codes, cached graph nodes, and navigation indexes. 
In particular, PQ codes grow with both dataset size and dimensionality, which largely explains the substantial memory gap—ranging from 3.4$\times$ on GloVe to 18$\times$ on OpenAI3072, compared with all-in-disk methods.

Regarding disk usage, Gorgeous consistently requires the most storage, reaching $30.708$ GB on OpenAI3072, due to its cache-oriented design with separate adjacency/vector files and redundant neighbor storage. In contrast, PageANN (all-in-disk) achieves the most compact indexes by avoiding duplicate neighbor storage and applying neighbor pruning, requiring only $0.514$ GB on GloVe and $0.558$ GB on SIFT1M. In high-dimensional settings, PipeANN exhibits the lowest disk usage among all methods (e.g., $15.259$ GB on the 3,072-dimensional OpenAI dataset) because it introduces less auxiliary-file overhead during search. 

\vspace{-0.05in}
\begin{tcolorbox}[
  colback=black!5,
  colframe=white,
  boxrule=0pt,
  arc=2pt,
  left=6pt,right=6pt,top=4pt,bottom=4pt
]
\textbf{Findings:} (1) Major-in-disk strategies require substantially more memory, whereas all-in-disk methods maintain a nearly constant footprint regardless of dataset scale.
(2) In low dimensions, disk usage is primarily determined by disk layout and graph structure. In high dimensions, it is dominated by raw vector size, resulting in similar usage across different methods.

\end{tcolorbox} 
\vspace{-0.1in}

\subsubsection{Layout Strategy Evaluation.}
\label{sec:layout_evaluation}
We evaluate global and local layout strategies separately to understand their individual impact on I/O behavior and query performance.

\begin{figure}[h]
    \centering
    \includegraphics[width=\linewidth]{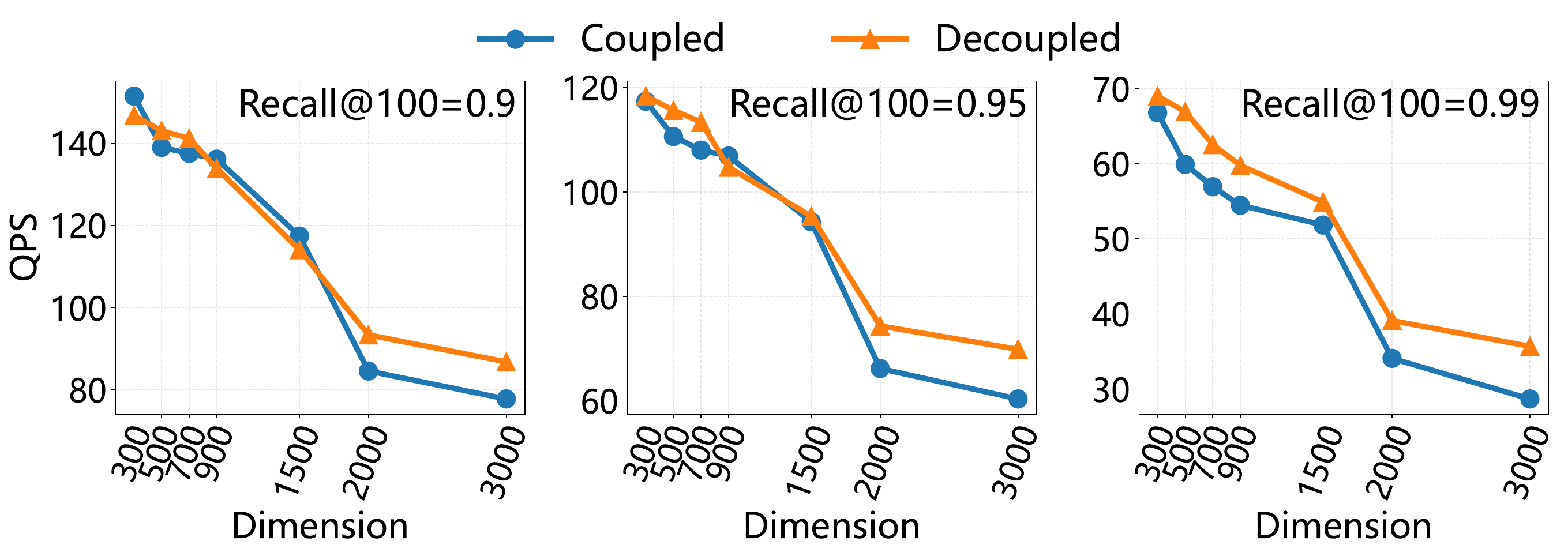}
    \vspace{-0.30in}
    \caption{QPS vs. dimension for different global layouts.}
    \vspace{-0.15in}
    \label{Fig:qps_dimension}
\end{figure}

\noindent
\textbf{Global Layout Evaluation.} 
Figure~\ref{Fig:qps_dimension} shows the QPS of two global layout strategies on OpenAI3072 across various dimensions and recall targets. Overall, the QPS of both strategies declines as dimensionality increases. Coupled layout performs better at lower dimensionalities ($\leq 1500D$), while decoupled layout becomes more effective at higher dimensionalities ($\geq 1500D$), especially under high recall targets.
Note that this $1500D$ threshold  is hardware-dependent. SSDs with faster random-read performance can reduce relative advantage of decoupled layout in accessing broader graph adjacency lists with fewer I/Os, shifting its favorable threshold over coupled layout toward higher dimensionalities.

Moreover, QPS decline is not smooth but exhibits several discrete drops. For example, at 0.9 recall, both layouts show only minor QPS reductions from 500D to 700D (2.05\% and 6.43\%, respectively), but experience a much sharper decline of about 14.5\% from 900D to 1500D. This behavior mainly arises because higher dimensionality often requires larger page sizes to store nodes, making QPS drop discretely when the page size increases.
Meanwhile, even under a fixed page size, coupled and decoupled layouts store different numbers of nodes per page, leading to different page  utilization and performance fluctuations. For instance, at 0.9 recall with a 4KB page size and 500D vectors, a coupled-layout page fits only one node (53.61\% page utilization), whereas the decoupled layout fits two nodes (97.66\%). Consequently, the decoupled layout outperforms the coupled layout at this specific lower dimensionality.

\vspace{-0.17in}

\vspace{-0.04in}
\begin{tcolorbox}[
  colback=black!5,
  colframe=white,
  boxrule=0pt,
  arc=2pt,
  left=6pt,right=6pt,top=4pt,bottom=4pt
]
\textbf{Findings:} 
(1) As dimensionality and recall increase, decoupled layouts become more advantageous by managing vectors and the graph index separately and effectively reducing vector reads (with a crossover around 1,500D).
(2) Higher dimensionality often necessitates larger page sizes, causing QPS to decline in discrete steps rather than smoothly.
(3) Even under the same dimensionality and page size, different global layouts lead to varying block utilization, affecting query performance.
\end{tcolorbox}

\begin{figure}[h]
    \centering
    \includegraphics[width=\linewidth]{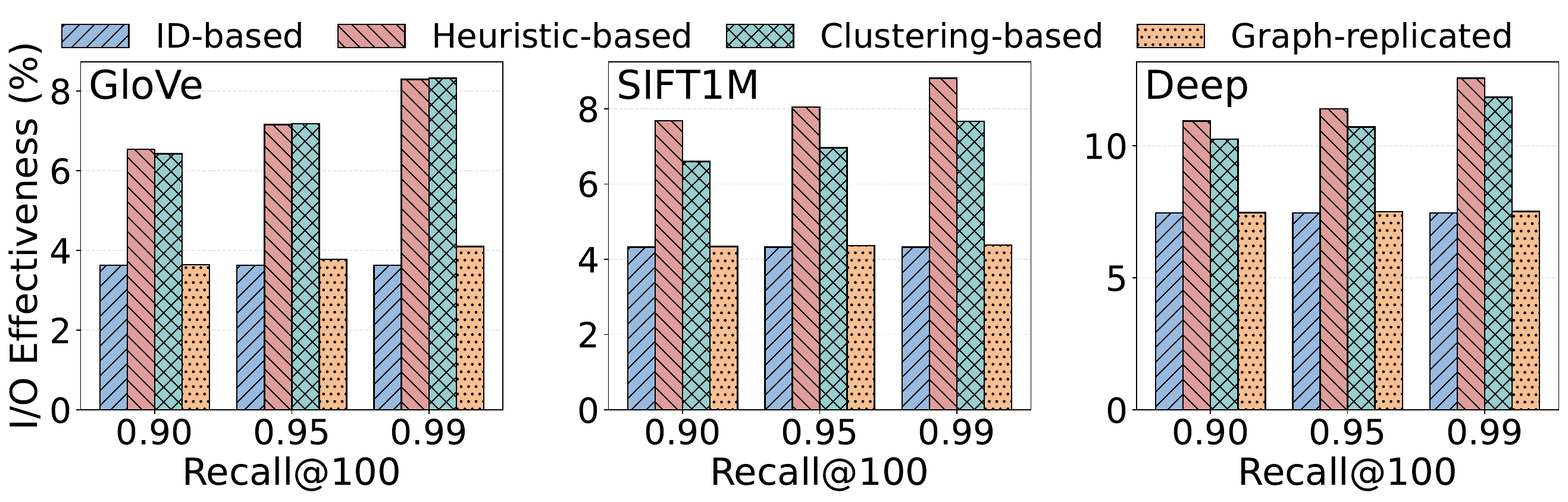}
    \vspace{-0.3in}
    \caption{I/O effectiveness of local layouts (16KB page size).}
    \vspace{-0.1in}
    \label{Fig:IO_Utilization}
\end{figure}

\noindent
\textbf{Local Layout Evaluation.}
To evaluate block layout effectiveness, we measure I/O effectiveness, defined as the ratio between the data actually required by the search path and the total data read under a fixed 16KB page size. We use the lower-dimensional GloVe, SIFT1M, and Deep datasets, where each block can accommodate more nodes and thus better expose layout differences.
Figure~\ref{Fig:IO_Utilization} shows that heuristic-based layouts consistently achieve the highest I/O effectiveness, outperforming ID-based, clustering-based, and graph-replicated layouts by 1.91$\times$, 1.16$\times$, and 1.83$\times$ on average, respectively. The advantage further increases at higher recall. For example, on GloVe, the gap between heuristic-based and ID-based layouts grows from 1.80$\times$ at 0.9 recall to 2.28$\times$ at 0.99 recall.
Despite these improvements, overall  effectiveness remains low: even the best case achieves only 12.55\% I/O effectiveness, indicating that most loaded data remains unused during I/O.

\vspace{-0.15in}

\vspace{-0.05in}
\begin{tcolorbox}[
  colback=black!5,
  colframe=white,
  boxrule=0pt,
  arc=2pt,
  left=6pt,right=6pt,top=4pt,bottom=4pt
]

\textbf{Findings:} 
(1) Overall I/O effectiveness remains \emph{low} across all methods, not exceeding 15\%.
(2) Heuristic-based layouts achieve higher I/O effectiveness by leveraging graph topology.
\end{tcolorbox} 
\vspace{-0.05in}

\vspace{-0.2in}
\begin{figure}[h]
    \centering
    \includegraphics[width=\linewidth]{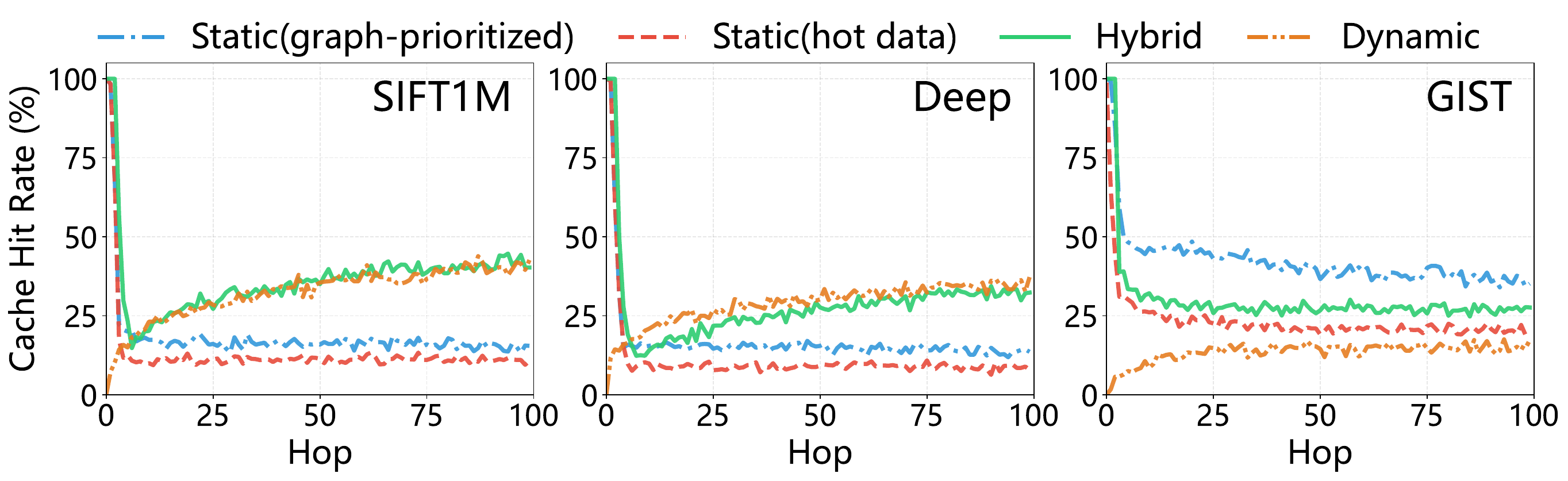}
    \vspace{-0.3in}
    \caption{Cache hit rate vs. search hops.}
    \vspace{-0.2in}
    \label{fig:cache_hit_rate}
\end{figure}

\subsubsection{Cache Strategy Evaluation}
\label{sec:cache_evaluation}
We measure caching effectiveness via the per-hop cache hit rate, i.e., the fraction of accesses served from cache. 
Figure~\ref{fig:cache_hit_rate} compares four strategies across datasets. Dynamic and hybrid strategies excel on low-dimensional datasets, outperforming static (graph-prioritized) baseline by 2.75$\times$ and 2.60$\times$ on SIFT1M, and 1.24$\times$ and 1.08$\times$ on Deep. They also surpass static (hot-data) strategy by up to 4.69$\times$ on Deep. However, on high-dimensional GIST dataset, static (graph-prioritized) achieves a 2.22$\times$ higher hit rate than dynamic baseline, which drops to 15.82\%. This shift arises from dimensionality: in low dimensions, dynamic/hybrid strategies fetch more useful vectors per I/O, boosting hits. As dimensionality grows, fewer vectors fit per block, reducing dynamic gains. Static (graph-prioritized) instead prioritizes compact adjacency lists, covering a broader graph topology within a 1\% memory budget, maintaining higher hit rates along long-tail search paths.

\vspace{-0.05in}
\begin{tcolorbox}[
  colback=black!5,
  colframe=white,
  boxrule=0pt,
  arc=2pt,
  left=6pt,right=6pt,top=4pt,bottom=4pt
]
\textbf{Findings:} 
The optimal caching strategy depends on dimensionality: \emph{dynamic} and \emph{hybrid} caching perform best in low dimensions, whereas \emph{static (graph-prioritized)} is more effective in high dimensions.
\end{tcolorbox} 
\vspace{-0.25in}
\begin{figure}[h]
    \centering
    \includegraphics[width=\linewidth]{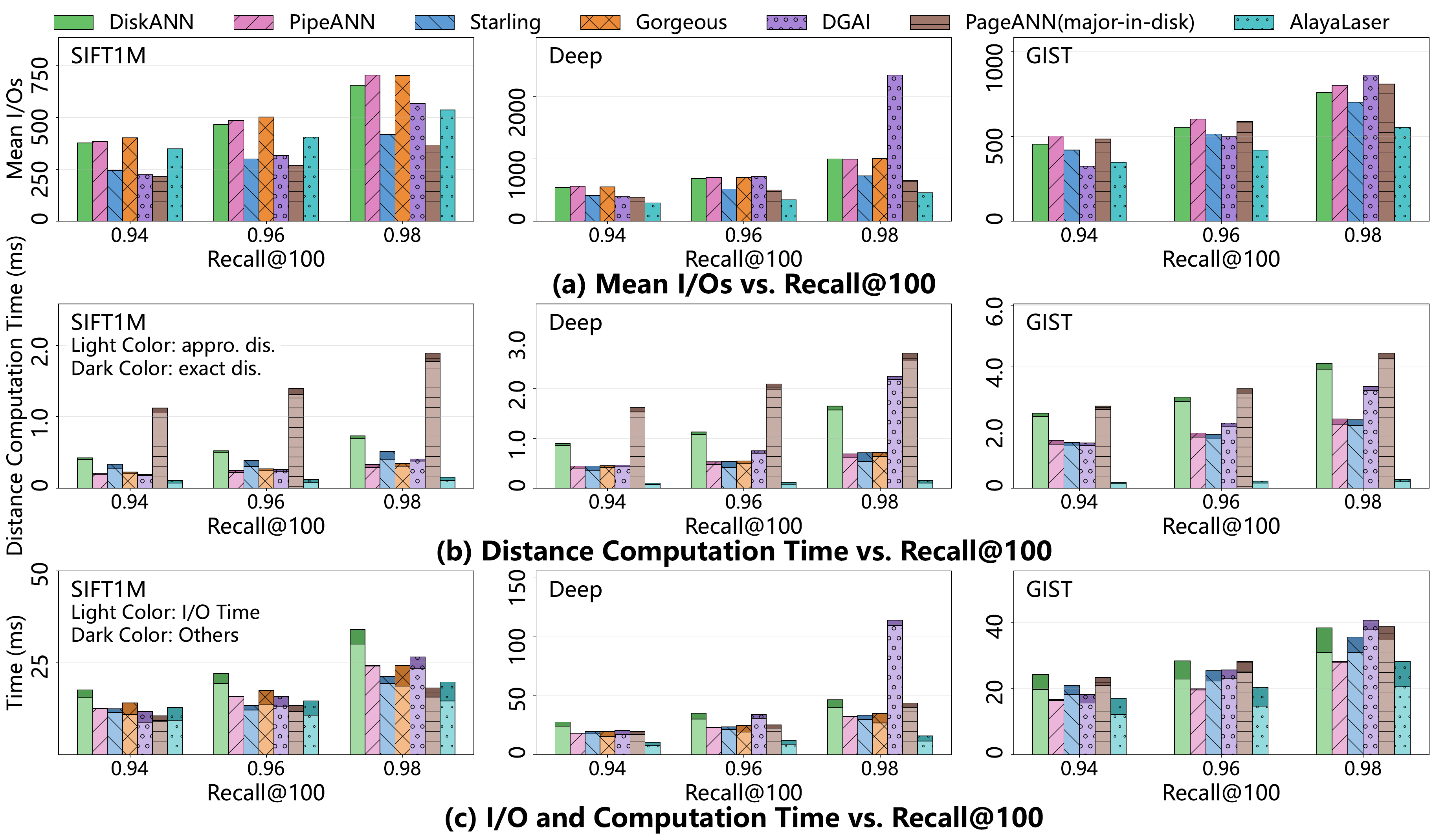}
    \vspace{-0.3in}
    \caption{Comparison of mean I/Os, computational overhead, and time decomposition for different search strategies.}
    \vspace{-0.2in}
    \label{fig:io_distance}
\end{figure}

\subsubsection{Query Strategy Evaluation}

\label{sec:query_evaluation}
Figures~\ref{fig:io_distance} (a)-(c) report mean I/Os, average distance-computation overhead, and the time breakdown between I/O and other stages across recall levels. All-in-disk methods except AlayaLaser are excluded due to their non-comparable I/O scales. The main observations are as follows. (1) Mean I/Os: PageANN achieves the fewest I/Os on SIFT1M, reducing mean I/Os by 11.81\%--47.13\% due to its effective layout strategy in low dimensions. On Deep and GIST, AlayaLaser achieves the lowest I/O cost, reducing mean I/Os by 29.24\%--68.25\% and 19.21\%--30.65\%, respectively, mainly because its combined PCA and RaBitQ approach introduces smaller approximation errors on these datasets than other PQ-based methods. (2) Computation: Approximate distance computation dominates computational overhead, while exact distance evaluation contributes only 2.89\%--24.64\%. AlayaLaser incurs significantly lower computation time due to its SIMD-friendly layout, whereas PageANN has the highest overhead, reaching 5.45$\times$--5.77$\times$ that of PipeANN on SIFT1M because of extensive PQ-based entry-point distance calculations. (3) I/O and computation time: Disk I/O time dominates the overall execution, accounting for 71.13\% to 99.77\% of the total query time. It is noteworthy that although AlayaLaser exhibits very low I/O and distance computation times, its end-to-end query time is only the best on Deep dataset. This happens because, on low-dimensional datasets like SIFT1M, AlayaLaser relies on an ID-based layout strategy, which provides poorer locality than the optimized layouts used by other methods. On high-dimensional datasets like GIST, AlayaLaser requires a larger page size to accommodate RaBitQ codes, resulting in a longer end-to-end latency. Moreover, PipeANN's I/O-driven overlapping is highly effective at hiding latency, making it slightly outperform AlayaLaser on GIST.

\vspace{-0.05in}
\begin{tcolorbox}[
  colback=black!5,
  colframe=white,
  boxrule=0pt,
  arc=2pt,
  left=6pt,right=6pt,top=4pt,bottom=4pt
]
\textbf{Findings:} 
(1) I/O-driven overlapping incurs no additional I/O under a consistent layout in a single-threaded setting.
(2) In low dimensions, optimized layouts (e.g., Starling and PageANN) substantially reduce I/O accesses.
(3) Asynchronous execution masks computation cost, making I/O dominant in latency. However, as dimensionality increases, computation cost becomes more significant.
\end{tcolorbox}
\vspace{-0.05in}

Notably, our conclusion that the search is I/O-bound for higher-dimensional datasets ($>512$) differs from the compute-bound observation in AlayaLaser~\cite{alayalaser}. This difference mainly comes from different parameter settings and hardware platforms. Applying the roofline model in \cite{alayalaser} to our GIST settings ($R=48$, 120 PQ subspaces, and 4KB pages) yields an operational intensity of $I \approx 1.41$ FLOP/byte, while our hardware has $I_{max} \approx 65.73$ FLOP/byte. Since $I < I_{max}$, our setup remains I/O-bound for higher-dimensional data.

\vspace{-0.1in}
\begin{figure}[h]
    \centering
    \includegraphics[width=\linewidth]{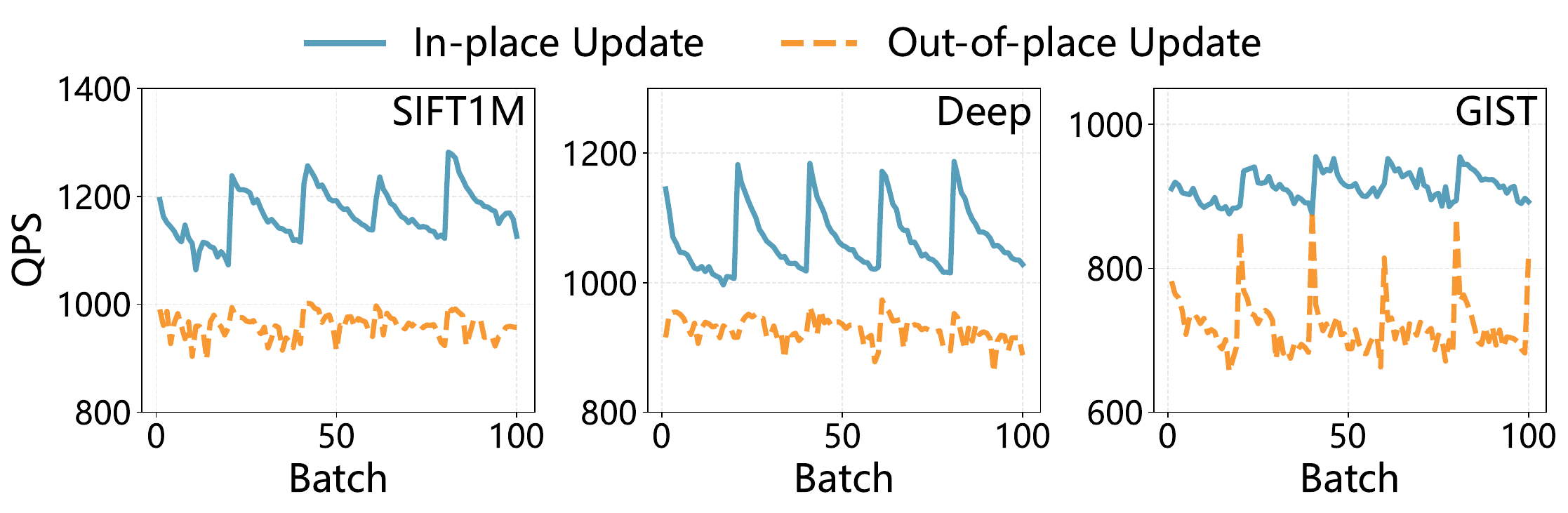}
    \vspace{-0.3in}
    \caption{Search QPS under concurrent updates.}
    \vspace{-0.15in}
    \label{fig:update_performance}
\end{figure}
\subsubsection{Update Efficiency}
\label{sec:update_evaluation}
\begin{figure}[h]
    \centering
    \begin{minipage}{0.8\linewidth}
        \centering
        \includegraphics[width=\linewidth]{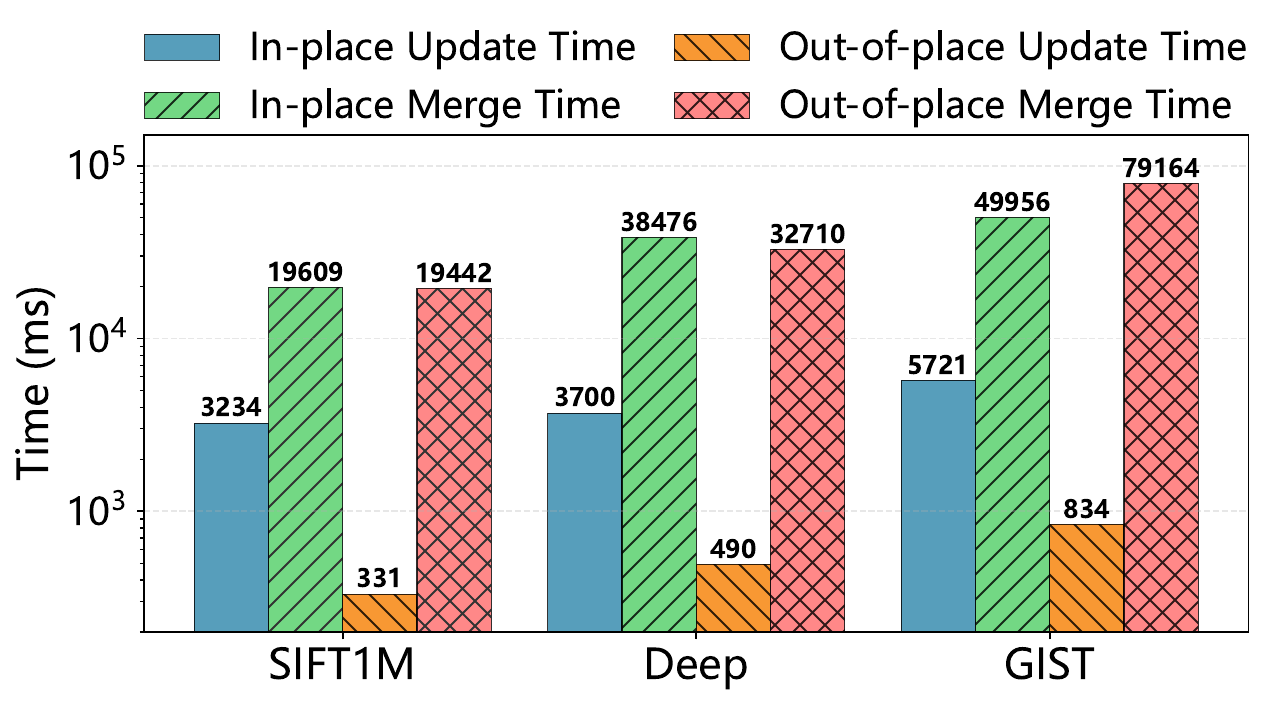}
    \end{minipage}
    \vspace{-0.2in}
    \caption{Average update and merge time.}
    \vspace{-0.2in}
    \label{fig:update_time_comparison}
\end{figure}

We evaluate in-place (OdinANN~\cite{ODINANN}) and out-of-place (FreshDiskANN~\cite{FreshDiskANN}) updates by executing 200K updates with a 1:1 insert-to-delete mix on an 800K-vector initial index. Following prior dynamic ANNS evaluations~\cite{FreshDiskANN,ODINANN,Quake}, this insert-to-delete mix simulates a realistic steady-state streaming-update environment where the overall index size remains stable. As shown in Figure \ref{fig:update_performance}, the in-place strategy consistently achieves higher sustained search throughput, outperforming the out-of-place strategy by approximately 1.22$\times$, 1.15$\times$, and 1.27$\times$ on SIFT1M, Deep, and GIST, respectively. Figure \ref{fig:update_time_comparison} compares update latency and merge time across strategies. The out-of-place strategy consistently achieves lower update latency than in-place. On SIFT1M, Deep, and GIST, out-of-place update latency is 9.7×, 7.5×, and 6.8× lower than in-place, respectively, and its merge times are 0.99×, 0.85×, and 1.58× those of the in-place.   

\vspace{-0.15in}

\begin{tcolorbox}[
  colback=black!5,
  colframe=white,
  boxrule=0pt,
  arc=2pt,
  left=6pt,right=6pt,top=4pt,bottom=4pt
]
\textbf{Findings:} 
The in-place update favors query-heavy workloads by sustaining higher QPS, whereas the out-of-place update better suits update-heavy or balanced workloads due to its lower update latency.
\end{tcolorbox}

\section{Lessons Learned}
\label{sec:lessons}

\subsection{Technique Selection}
\label{sec:tech-selection}
\begin{figure}[t]
    \centering
    \includegraphics[width=\linewidth]{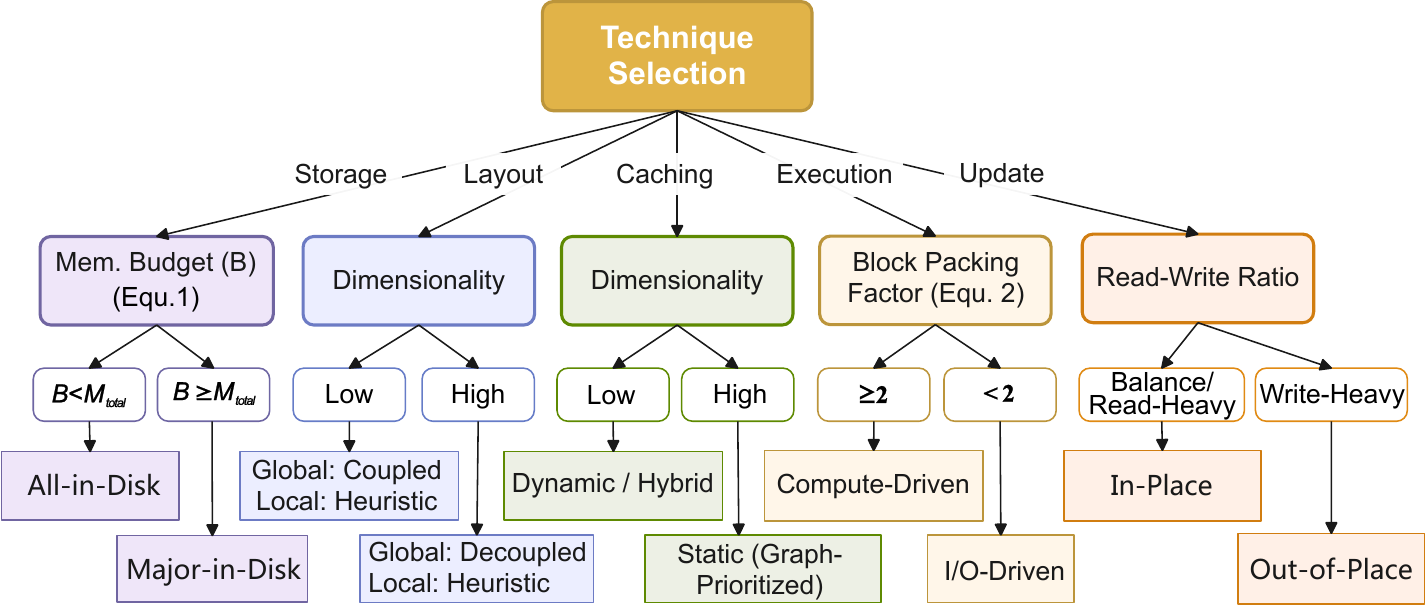}
    \vspace{-0.2in}
    \caption{Technique selection.}
    \vspace{-0.1in}
    \label{fig:technique_selection}
\end{figure}

Based on the experimental analysis, we summarize a decision tree (Figure~\ref{fig:technique_selection}) to guide technique selection across five key dimensions.

\noindent
\textbf{(1) Storage:}
Under a memory budget $B$, we estimate the total memory footprint as
\begin{equation}
\label{eq:memory_total_new}
M_{\text{total}} =
M_{\text{nav}} + M_{\text{pq}} + M_{\text{cache}} + C ,
\end{equation}
where $M_{\text{nav}} = (d s_d + (1+R)s_i)N$, $M_{\text{pq}} = d s_d\,2^{8k} + n_{\text{pq}}kN$, and $M_{\text{cache}} = M_{\text{dynamic}}
+ N_{\text{hot}}(d s_d + (1+R)s_i)$.
Here, 
$N$ and $N_{\text{hot}}$ denote the numbers of total and cached hot vectors, respectively; $s_d$ and $s_i$ are the byte sizes of vector values and identifiers; $k$ and $n_{\text{pq}}$ denote the PQ code size and number of PQ chunks; $M_{\text{dynamic}}$ is the dynamic cache space; and $C$ captures working memory and auxiliary structures (typically $\leq 50$MB). If $M_{\text{total}}\le B$, we adopt the \emph{major-in-disk} strategy; otherwise, we switch to \emph{all-in-disk}.



\noindent
\textbf{(2) Layout:} Layout is primarily determined by vector dimensionality, which directly affects I/O efficiency. We use \emph{coupled global layout} for low-dimensional data ($d \le 1500$) and \emph{decoupled global layout} otherwise. For local layout, \emph{heuristic-based} strategy is adopted as a practical baseline, while more advanced designs remain an open direction (Section~\ref{sec:further_direction}).

\noindent
\textbf{(3) Caching:} The caching strategy is determined by vector dimensionality.
For low-dimensional data (typically $d \le 900$), \emph{dynamic} or \emph{hybrid} caching is preferred.
For high-dimensional data, \emph{static (graph-prioritized)} caching is recommended.

\noindent
\textbf{(4) Query Execution:} The strategy relies on the block packing factor (BPF), defined as
\begin{equation}
\text{BPF} = \frac{\text{page\_size}}{d\,s_d + (1+R)s_i},
\end{equation}
which measures how many vectors fit in a disk page. If $\text{BPF}<2$, I/O dominates and we adopt an \emph{I/O-driven} strategy with I/O--compute overlapping. If $\text{BPF}\ge 2$, multiple vectors amortize per page, making computation dominant and a \emph{compute-driven} strategy preferable.


\noindent
\textbf{(5) Update:} The update technique is based on read–write ratio. For balanced or read-heavy workloads, in-place update technique is preferred.
For write-heavy workloads, out-of-place is recommended.

\vspace{0.05in}
\subsection{Further Directions}
\label{sec:further_direction}
Based on our findings and analysis, several open problems and research challenges remain in disk-resident graph-based ANN search.

\noindent
\textbf{(1) Adaptive Design under Varying Dimensionality.}
We observe that several methods exhibit markedly different performance as dimensionality increases due to degraded block locality and the higher cost of retrieving raw vectors. This motivates \emph{dimension-aware designs} as future work.

\noindent
\textbf{(2) More Effective Layouts for Higher I/O Effectiveness.}
Our results show that overall I/O effectiveness remains notably low across methods, primarily since current layouts fail to colocate consecutive nodes along a search path into the same disk page.
This motivates the design of more effective layout strategies that explicitly consider locality in graph traversals and the corresponding access patterns.
Future directions include exploring traversal-aware or workload-adaptive layouts that dynamically reorganize disk blocks to enhance spatial locality and raise effective utilization.

\noindent
\textbf{(3) More Efficient All-in-Disk Designs.}
While all-in-disk designs offer attractive memory efficiency, our results show they generally lag behind major-in-disk methods in query performance. This gap is largely due to frequent disk accesses and limited opportunities for I/O–compute overlap. Designing more efficient all-in-disk methods that can better exploit locality, caching, and asynchronous execution, while preserving their low memory footprint, remains an important and open research direction.

\noindent
\textbf{(4) Parameter Tuning under System and Workload Diversity.}
The performance of disk-resident ANN systems is highly sensitive to a set of parameters, such as page size and beam width.
Our findings indicate that changes in dimensionality can trigger stepwise performance degradation due to page-size shifts, while different execution strategies favor different parameter settings.
Automatically tuning these parameters under varying datasets, dimensionalities, and hardware configurations remains an open challenge.

\noindent
\textbf{(5) Extension to Cloud-Native Multi-Layer Storage.} 
Most existing disk-resident ANN systems are designed for heterogeneous memory-disk storage. However, modern cloud environments increasingly adopt multi-layer storage hierarchies, including memory, local SSD, and object storage. Extending disk-resident ANN designs to cloud-native, multi-layer storage settings raises new challenges in data placement, caching, and execution scheduling, especially under variable latency and bandwidth constraints.

\vspace{0.1in}
\section{Conclusion}
In this paper, we systematically analyze disk-resident graph-based ANN methods across five dimensions: storage, layout, caching, query execution, and updates. Through extensive evaluations, we identify the key trade-offs underlying performance differences and distill a set of practical findings. Based on these insights, we present a decision guide to help select suitable designs for scalable ANN search and highlight promising future research directions.


\begin{acks}
\begin{sloppypar}

This paper was supported in part by Fundamental and Interdisciplinary Disciplines Breakthrough Plan of the Ministry of Education of China (No. JYB2025XDXM103), New Generation Artificial Intelligence-National Science and Technology Major Project (No. 2026ZD0128803), National Key R\&D Program of China (No. 2023YFB4503600), NSF of China (62432007, 62441236, 62372296, 62525202, 62232009, 62502304, U25A20437), Hong Kong Research Grants Council Grant (No. 12202024), Huawei, Tencent, and ByteDance.

\end{sloppypar}
\end{acks}

\clearpage

\balance
\bibliographystyle{ACM-Reference-Format}
\bibliography{reference}
\balance

\end{document}